\documentclass{statsoc}
\usepackage{float}
\usepackage{graphics}
\usepackage{epsfig}
\usepackage{amssymb}
\usepackage{amsmath}
\usepackage{setspace}
\usepackage{anysize}

\def\be{\begin{equation}}
\def\ee{\end{equation}}
\def\ba{\begin{array}}

\def\ea{\end{array}}

\newtheorem{theorem}{Theorem}

\newtheorem{corollary}{Corollary}

\newbox\TempBox\newbox\TempBoxA
\newcommand{\beq}{\begin{equation}}
\newcommand{\eeq}{\end{equation}}
\marginsize{3cm}{3cm}{3cm}{3cm}

\title[Hourly ozone fields]{Modeling Hourly Ozone Concentration
Fields}

\author[Yiping Dou {\it et al.}]{Yiping Dou}
\address{University of British Columbia,
Vancouver,
Canada.}
\email{ydou@stat.ubc.ca}
\author{Nhu D Le}
\address{BC Cancer Research Centre,
Vancouver,
Canada}
\author{James V Zidek}
\address{University of British Columbia,
Vancouver,
Canada.}

\begin{document}
\begin{abstract}This paper presents a dynamic linear model for modeling
hourly ozone concentrations over the eastern United States. That
model, which is developed within an Bayesian hierarchical framework,
inherits the important feature of such models that its coefficients,
treated as states of the process, can change with time. Thus the
model includes a time--varying site invariant mean field as well as
time varying coefficients for 24 and 12 diurnal cycle components.
This cost of this model's great flexibility comes at the cost of
computational complexity, forcing us to use an MCMC approach and to
restrict application of our model domain to a small number of
monitoring sites. We critically assess this model and discover some
of its weaknesses in this type of application.
\end{abstract}

\section{Introduction}\label{intro:DLM}

This paper presents a model for the spatio--temporal field of hourly
ozone concentrations for subregions of the eastern United States,
one that can in principle be used for both spatial and temporal
prediction. It goes on to critically assess that model and the
approach used for its construction, with mixed results.

Such models are needed for a variety of purposes described in Ozone
(2005) where a comprehensive survey of the literature on such
methods is given, along with their strengths and weaknesses. In
particular, they can be used to help characterize population levels
of exposures to ozone in outdoor environments, based on measurements
taken at often remote ambient monitors.

These interpolated concentrations can also be used as input to
computer models that incorporate indoor environments to more
accurately predict population levels of exposure to an air
pollutant. Such models can reduce the deleterious effects of errors
resulting from the use of ambient monitoring measurements to
represent exposure. For example, on hot summer days the ambient
levels will overestimate exposure since people tend to stay in air
conditioned indoor environments where exposures are lower. To
address that problem, the US Environmental Protection Agency (EPA)
developed APEX. It is being used by policy--makers to set air
quality standards under hypothetical emission reduction scenarios
(Ozone, 2005). Interpolated ozone fields could well be used as input
to APEX to further reduce that measurement error although that
application has not been made to date for ozone. However, it has
been made for particulate air pollution through an exposure model
called SHEDS (Burke et al., 2001) as well as a simplified version of
SHEDS (Calder et al., 2003).

Interest in predicting human exposure and hence in mapping ozone
space--time fields, stems from concern about the adverse human
health effects of ozone. Ozone (2005) reviews an extensive
literature on that subject. Exposure chamber studies show that
inhaling high concentrations of ozone compromises lung function
quite dramatically in healthy individuals (and presumably to an even
greater degree in unhealthy individuals such as those suffering from
asthma). Moreover, epidemiology studies show strong associations
between adverse health effects such exposures. Consequently, the US
Clean Air Act mandates that National Ambient Air Quality Standards
are necessary for ozone to protect human health and welfare. Thus,
spatio--temporal models can have a role in setting these NAAQS.

Ozone concentrations over a geographic region vary randomly over
time, and therefore constitute a spatio--temporal field. In both
rural and urban areas such fields are customarily monitored, the
latter to ensure compliance with the NAAQS amongst other things. In
fact, failure can result in substantial financial penalties.

A number of approaches can be taken to modelling such space time
fields. Here we investigate a promising one that involves selecting
a member of a very large class of so--called state space models.
Section \ref{chapter2:dlmbackgrand} describes our choice, a dynamic
linear model (DLM), a variation of those proposed by Huerta et al.
(2004) and Stroud et al. (2001). Here ``dynamic'', refers to the
DLM's capability of systematically modifying its parameters over
time, a seemingly attractive feature since the processes it models
will themselves generally evolve ``due to the passage of time as a
fundamental motive force'' (West and Harrison, 1997). However, other
approaches are possible and in a companion report currently in
preparation, the DLM selected here will be compared with other
possibilities.

Section \ref{chapter2:dlmbackgrand} introduces the hourly
concentration field that is to be modeled in this report. There
consideration of measurements made at fixed site monitors and
reported in the AIRS dataset leads to the construction of our DLM.
[The {\it EPA} (Environmental Protection Agency) changed the {\it
AIRS} (Aerometric Information Retrieved System) to the {\it AFS}
(Air Facility Subsystem) in 2001.] That model becomes the object of
our assessment in subsequent sections. To illustrate how to select
some of the model parameters in the DLM, we use the simple
first--order polynomial DLM in Section \ref{pred:var:simplest:dlm}
to shed some light on this problem. Moreover, we prove there in a
simple but representative case, that under the type of model
constructed here and by Huerta et al. (2004), the predictive
variances for successive time points conditional on all the data
must be monotonically increasing, an undesirable property.
Theoretical results and algorithms on the DLM are represented in
Sections \ref{chapter2:algo:est} and \ref{chapter2:algo:pred}. The
MCMC sampling scheme is outlined in Section
\ref{chapter2:algo:est:mcmc}. The
forward--filtering--backward--sampling (FFBS) method is demonstrated
in Section \ref{ffbs:state:sample} to estimate the state parameters
in the DLM. Moreover, we outline the MCMC sampling scheme to obtain
samples for other model parameters from their posterior conditional
distributions with a Metropolis--Hasting step. Section
\ref{chapter2:algo:pred} gives theoretical results for prediction
and interpolation at unmonitored (ungauged) sites from their
predictive posterior distributions. Section
\ref{chapter2:example:Cluster2} shows the results of MCMC sampling
along with interpolation results on the ozone study. Section
\ref{problems:in:DLM} describes problems with the DLM process
revealed by our assessment. We summarize our findings and draw
conclusions from our assessment in Section \ref{chapter2:summary}.

As an added note, we have developed software, written in C and R and
available online (http://enviro.stat.ubc.ca) that may be used to
reproduce our findings or to use the model for modeling hourly
pollution in other settings.

\section{Model development}\label{chapter2:dlmbackgrand}

Although we believe the methods described in this paper apply quite
generally to hourly pollution concentration space--time fields, it
focuses on an hourly ozone concentrations (ppb) over part of the
eastern United States during the summer of 1995. In all, $375$
irregularly located sites (or ``stations'') monitor that field. To
enable a focused assessment of the DLM approach and make
computations feasible, we consider just one cluster of ten stations
(Cluster 2), in close proximity to one another. However, in work not
reported here for brevity, two other such clusters led to similar
findings. Note that Cluster 2 has the same number of stations as the
one in Mexico City studied by Huerta et al.(2004).

The initial exploratory data analysis followed that of Huerta et al.
(2004) with a similar result, a square--root transformation of the
data is needed to validate the normality assumption for the DLM
residuals. [Note that a small amount of randomly missing data were
filled in by the spatial regression method (SRM), before we began.]
The plot of a Bayesian periodogram (Dou et al., 2007) for the
transformed data at the sites in our cluster reveals a peak between
1 pm and 3 pm each day with a significant $24$--hour cycle for the
stations in Cluster 2. We also found a slightly significant
$12$--hour cycle. However, no obvious weekly cycles or nightly peaks
were seen. Thus, the DLM suggested by our analysis turns out to be a
variant of the one in Huerta et al. (2004); it has states for both
local trends as well as periodicity across sites.

To define the model, let $y_{it}$ denote the square--root of the
observable ozone concentration, at site $\mathbf{s_{i}},$
$i=1,\ldots,n,$ and time $t,$ $t=1,\ldots,T,$ $n$ being the total
number of gauged (that is, monitoring) sites in the geographical
subregion of interest and $T,$ the total number of time points.
Furthermore, let ${\mathbf{y_{t}}} = (y_{1t},\ldots,
y_{nt})^{\prime} : n \times 1$. Then the DLM for the field is
\begin{eqnarray}
{\mathbf{y_{t}}} & = & {\mathbf{1}}_{n}^{\prime}\beta_{t} +
S_{1t}(a_{1}){\mathbf{\alpha_{1t}}} +
S_{2t}(a_{2}){\mathbf{\alpha_{2t}}} + {\mathbf{\nu_{t}}}
\label{dlm:y} \\
\beta_{t}        & = & \beta_{t-1} + w_{t} \label{dlm:beta}\\
{\mathbf{\alpha_{jt}}} & = & {\mathbf{\alpha_{j,t-1}}} +
{\mathbf{\omega_{t}}}^{\alpha_{j}}, \label{dlm:alpha}
\end{eqnarray}
where ${\mathbf{\nu_{t}}} \sim N[{\mathbf{0}},
\sigma^{2}V_{\lambda}],$ $w_{t} \sim N[0, \sigma^{2}\tau_{y}^{2}],$
${\mathbf{\omega_{t}}}^{\alpha_{j}} \sim N[{\mathbf{0}},
\sigma^{2}\tau_{j}^{2}V_{\lambda_{j}}],$
$V_{\lambda}=\exp(-V/\lambda),$ $V_{\lambda_{j}} =
\exp(-V/\lambda_{j})$, $j=1,2,$ and ${\mathbf{\alpha_{jt}}} =
(\alpha_{j1t},\ldots,\alpha_{jnt})^{\prime}: n \times 1, j=1,2.$
Here $\beta_{t}$ denotes a canonical spatial trend and
$\alpha_{jit},$ a seasonal coefficient for site $i$ at time $t$
corresponding to a periodic component, $S_{jt}(a_{j}),$ where
$S_{jt}(a_{j}) = \cos(\pi tj/12) + a_{j}\sin(\pi tj/12), j=1,2.$
Note that $V=(v_{ij}): n \times n$ represents the distance matrix
for the gauged sites ${\mathbf{s_{1}}}, \ldots, {\mathbf{s_{n}}},$
that is, $v_{ij}=||{\mathbf{s_{i}}} - {\mathbf{s_{j}}}||$ for
$i,j=1,\ldots,n,$ where $||{\mathbf{s_{i}}} - {\mathbf{s_{j}}}||$
denotes the Euclidean distance (km) between sites ${\mathbf{s_{i}}}$
and ${\mathbf{s_{j}}}.$

Models (\ref{dlm:y})--(\ref{dlm:alpha}) can also written in the form
of a state space model with the observation and state equations
\begin{eqnarray}\label{dlm:obs}
\mathbf{y_{t}} & = & \mathbf{F_{t}^{\prime}}\mathbf{x_{t}} +
\mathbf{\nu_{t}}
\end{eqnarray}
\vspace*{-0.9cm}
\begin{eqnarray}\label{dlm:state}
\mathbf{x_{t}} & = & \mathbf{x_{t-1}} + \mathbf{\omega_{t}},
\end{eqnarray}
where $\mathbf{x_{t}^{\prime}} = (\beta_{t},
{\mathbf{\alpha_{1t}}}^{\prime}, {\mathbf{\alpha_{2t}}}^{\prime}),$
${\mathbf{\omega_{t}}}^{\prime}= (\omega_{t},
{{\mathbf{\omega_{t}}}^{\alpha_{1}}}^{\prime},
{{\mathbf{\omega_{t}}}^{\alpha_{2}}}^{\prime})^{\prime},$ and
$\mathbf{F_{t}^{\prime}}$ is given by
$$
\left[
  \begin{array}{ccccccccc}
  1 & S_{1t}(a_{1}) & 0             & \ldots & 0             & S_{2t}(a_{2}) & 0             & \ldots  & 0 \\

  1 & 0             & S_{1t}(a_{1}) & \ldots & 0             & 0             & S_{2t}(a_{2}) & \ldots  & 0 \\

  \vdots            & \vdots        & \vdots &               & \vdots        & \vdots        & \vdots  &     & \vdots\\

  1 & 0             & 0             & \ldots & S_{1t}(a_{1}) & 0             & 0             & \ldots  & S_{2t}(a_{2})
  \end{array}
\right].
$$
Note that ${\mathbf{\omega_{t}}} \sim N[{\bf{0}}, \sigma^{2}W],$
$W$ being the block diagonal matrix with diagonal entries
$\tau_{y}^{2},$ $\tau_{1}^{2}\exp(-V/\lambda_{1}),$ and
$\tau_{1}^{2}\exp(-V/\lambda_{2}).$

Let $y_{1:T} = (y_{1:T}^{m}, y_{1:T}^{o})^{\prime},$ where
$y_{1:T}^{m} = (y_{1}^{m}, \ldots, y_{T}^{m})$ represents all the
missing values and $y_{1:T}^{o},$  all the observed values in
Cluster 2 sites for $t=1,\ldots,T.$ The model unknowns are therefore
the coordinates of the vector $(\lambda, \sigma^{2}, x_{1:T},
y_{1:T}^{m}, a_{1}, a_{2}),$ in which the vector of state parameters
up to time $T$ is $x_{1:T}=({\mathbf{x_{1}}},
\ldots,{\mathbf{x_{T}}})$, the range parameter is $\lambda$, the
variance parameter is $\sigma^{2}$  and finally the vector of phase
parameters is ${\mathbf{a}} = (a_{1}, a_{2})$. Let
${\mathbf{\gamma}} = (\tau_{y}^{2}, \tau_{1}^{2}, \lambda_{1},
\tau_{2}^{2}, \lambda_{2})$ be the vector of parameters fixed in the
DLM to render computation feasible.

Specification of the DLM is completed by prescribing the hyperpriors
for the distributions of some of the model parameters:
\begin{eqnarray*}
\lambda      & \sim & IG(\alpha_{\lambda}, \beta_{\lambda})\\
\sigma^{2}   & \sim & IG(\alpha_{\sigma^{2}}, \beta_{\sigma^{2}})\\
{\mathbf{a}} & \sim &  N({\mathbf{\mu}}_{a}^{o}, \Sigma_{a}^{o} ).
\end{eqnarray*}
Notice that $\lambda$ and $\sigma^{2}$ have inverse Gamma
distributions for computational convenience.\footnote[2]{$X \sim
IG(\alpha, \beta)$ if $Y=1/X \sim G(\alpha, \beta),$ where $p(y)
\propto y^{\alpha-1}\exp(-\beta y)$ for $\alpha,\beta>0.$} The
choice of the hyperpriors is discussed in Section
\ref{chapter2:example:Cluster2}.

We express the state--space model in two different ways because of
our dual objectives of parameter inference and interpolation. For
simplicity, we use models (\ref{dlm:obs})--(\ref{dlm:state}) for
inference about the range, variance and state parameters (see
Section \ref{ffbs:state:sample}), and use models
(\ref{dlm:y})--(\ref{dlm:alpha}) for inference on the phase
parameters (see Section \ref{ffbs:phase:sample}) and interpolation
(see Section \ref{chapter2:algo:pred}).

\section{Parameter specification}\label{pred:var:simplest:dlm}

Before turning to the implementation of the approach in the next
section, we explore theoretically, albeit in a tractable special
case, some features of the model. That exploration leads to insight
about how the model's parameters should be specified as well as
undesirable consequences of inappropriate choices. Our assessment
will focus on the accuracy of the model's predictions.

This simple model we consider is a special case of the so--called
``first--order polynomial model'', a mathematically tractable,
commonly used model. It captures many important features and
properties of the DLM we have adopted.

For $i=0,1,\ldots,n$ and $t=1,\ldots,T$, the first--order polynomial
DLM is given by
\begin{eqnarray}
y_{it}    & = & \beta_{t} + \varepsilon_{it} \label{pred:var:obs:eqn}\\
\beta_{t} & = & \beta_{t-1} + \delta_{t}, \label{pred:var:state:eqn}
\end{eqnarray}
where ${\mathbf{\varepsilon_{t}}}=(\varepsilon_{0t}, \ldots,
\varepsilon_{nt})^{\prime} \sim N({\mathbf{0}},
\sigma_{\varepsilon}^{2} \exp(-V/\lambda)),$ and $\delta_{t} \sim
N(0, \sigma_{\delta}^{2}).$ Assume $\beta_{0} \sim
N(0,\sigma_{\beta}^{2})$ and $\lambda, \sigma_{\varepsilon}^{2},
\sigma_{\delta}^{2}$ and $\sigma_{\beta}^{2}$ are all currently known.

The first--order polynomial DLM is particularly useful for
short--term prediction since then the underlying evolution
$\beta_{t}$ is roughly constant. Observe that the zero--mean evolution
error $\delta_{t}$ process is independent over time, so that the
underlying process is merely a random walk; the model does not
anticipate long--term variation. At any fixed time $t:$
\begin{eqnarray}
\beta_{t} & = & \beta_{0} + \sum_{k=1}^{t}\delta_{k}
\label{pred:var:beta:eqn}\\
y_{it}    & = & \beta_{0} + \sum_{k=1}^{t}\delta_{k} + \varepsilon_{ik}
\label{pred:var:yit:eqn}.
\end{eqnarray}
Consequently, the first--order polynomial DLM has the following
covariance structure:
\begin{eqnarray}
\mbox{Var($y_{it}$)} & = & \sigma_{\beta}^{2} + t\sigma_{\delta}^{2}
+ \sigma_{\varepsilon}^{2}
\label{pred:var:var:yit} \\
\mbox{Cov($y_{it},  y_{jt}$)} & = & \sigma_{\beta}^{2} +
t\sigma_{\delta}^{2} + \sigma_{\varepsilon}^{2}\exp(-d_{ij}/\lambda)
\quad (i\neq j)
\label{pred:var:cov:yit:yjt}\\
\mbox{Cov($y_{it}, y_{js}$)} & = & \sigma_{\beta}^{2} + \min\{t,s\}
\sigma_{\delta}^{2}  \quad (s\neq t) , \label{pred:var:cov:yit:yjs}
\end{eqnarray}
where  $d_{ij} = || {\mathbf{s_{i}}} - {\mathbf{s_{j}}} ||,$ for
$i,j=0,1,\ldots,n$ and $t,s = 1,\ldots, T.$

This DLM defines a non--stationary spatio--temporal process since
for the first--order polynomial model to be stationary, the
eigenvalues of state transfer matrix,  $G=G_{t}$ in the notation of
West and Harrison (1997), must lie inside of the unit circle.
However, $G_{t}=1$ so that this process is not a stationary Gaussian
DLM. Furthermore, the DLM defined in Section
\ref{chapter2:dlmbackgrand} is non--stationary because $G_{t} =
I_{2n+1}$ given all the model parameters in
(\ref{dlm:obs})--(\ref{dlm:state}). The DLM in
(\ref{pred:var:obs:eqn})--(\ref{pred:var:state:eqn}) has an
important property that the covariance functions in
(\ref{pred:var:cov:yit:yjt})--(\ref{pred:var:cov:yit:yjs}) depends
on the time point of $\min\{t,s\}$, not on $|t-s|$ thus confirming
our observation of non-stationary.

We readily find the correlation between $y_{it}$ and $y_{js}$
to be
\begin{eqnarray}
\mbox{Cor($y_{it}, y_{jt}$)} & = & \frac{ \sigma_{\beta}^{2} +
t\sigma_{\delta}^{2} +
\sigma_{\varepsilon}^{2}\exp(-d_{ij}/\lambda)}{ \sigma_{\beta}^{2} +
t\sigma_{\delta}^{2} + \sigma_{\varepsilon}^{2} } \quad (i \neq j)
\label{pred:var:corr:yit:yjt}\\
\mbox{Cor($y_{it}, y_{js}$)} & = & \frac{ \sigma_{\beta}^{2} +
\min\{t,s\} \sigma_{\delta}^{2} }{ \sqrt{ \sigma_{\beta}^{2} +
t\sigma_{\delta}^{2} + \sigma_{\varepsilon}^{2} } \sqrt{
\sigma_{\beta}^{2} + s\sigma_{\delta}^{2} + \sigma_{\varepsilon}^{2}
}  } \quad (s\neq t) \label{pred:var:corr:yit:yjs}
\end{eqnarray}
where $i,j=0,\ldots,n$ and $s,t=1,\ldots,T.$
\vskip .1in
\noindent{\bf Remarks.}
\begin{description}
\item{{\bf 1.}} The correlations in (\ref{pred:var:corr:yit:yjt})
and (\ref{pred:var:corr:yit:yjs}) have the following properties
when $i\neq j$:
\begin{enumerate}
\item[(i)]
  \begin{eqnarray}\label{pred:var:corr:prop1}
  \mbox{Cor($y_{it}, y_{jt}$)} & > & \mbox{Cor($y_{it}, y_{js}$)}
  \end{eqnarray}
for $s\neq t, s,t=1,\ldots,T$ and
\item[(ii)]
  \begin{eqnarray}\label{pred:var:corr:prop2}
  \mbox{Cor($y_{it}, y_{jt}$)} - \mbox{Cor($y_{it}, y_{js}$)}
  \end{eqnarray}
is a monotone increasing function of $|t-s|$.
\end{enumerate}
Thus for any fixed time point $t$, $\mbox{Cor($y_{it}, y_{js}$)}$ as
a function of $s$ attains its maximum at $s=t$ and decreases as
$|s-t|$ increases.
\item{{\bf 2.}} By (\ref{pred:var:corr:yit:yjt}), $\mbox{Cor($y_{it}, y_{jt}$)}
\rightarrow 1$ as $t\rightarrow \infty$ for $i\neq j,$
$i,j\in\{0,\ldots,n\}.$ That property seems unreasonable; the degree
of association between two fixed monitors should not increase as an
artifact of a larger time t. That suggests a need to make some of
the model parameters, say $\sigma_{\delta}^{2}$, depend on time.
More specifically, (\ref{pred:var:corr:yit:yjt}) suggests making
$t\sigma_{\delta}^{2} = O(1)$ stabilize $\mbox{Cor($y_{it},
y_{jt}$)}.$
Carrying this assessment further, for any two sites in close
proximity, i.e. for $d_{ij}\simeq 0$,
$$ \mbox{Cor($y_{it},
y_{jt}$)} \simeq \frac{\sigma_{\beta}^{2} + t\sigma_{\delta}^{2}+
\sigma_{\varepsilon}^{2}}{ \sigma_{\beta}^{2} + t\sigma_{\delta}^{2}
+ \sigma_{\varepsilon}^{2} } = 1,
$$
a result that seems quite reasonable. For two sites very far apart
so that $d_{ij} \rightarrow \infty$,
$$ \mbox{Cor($y_{it}, y_{jt}$)} \rightarrow \frac{
\sigma_{\beta}^{2} + t\sigma_{\delta}^{2} }{ \sigma_{\beta}^{2} +
t\sigma_{\delta}^{2} + \sigma_{\varepsilon}^{2}} = \frac{
\sigma_{\beta}^{2} + O(1) }{ \sigma_{\beta}^{2} + O(1) +
\sigma_{\varepsilon}^{2} }.
$$
This correlation should be close to $0$. In other words, we should
have $\sigma_{\beta}^{2} + O(1) \ll \sigma_{\varepsilon}^{2}.$ A
sufficient condition for this to hold is $\sigma_{\beta}^{2} \ll
\sigma_{\varepsilon}^{2}$ and $t\sigma_{\delta}^{2} = O(1) \ll
\sigma_{\varepsilon}^{2}.$
\end{description}
The key result, (\ref{pred:var:corr:yit:yjt}), suggests a simple but
straightforward way to adjust the model parameter
$\sigma_{\delta}^{2}$ according to the size of $T$, namely, to
replace it by $\sigma_{\delta}^{2}/T$. That choice is empirically
validated in Section \ref{chapter2:summary}.

We turn now to study the behavior of the predictive variances in the
first--order polynomial DLM that helps us understand the
interpolation results. To that end consider the correlations of
responses at an ungauged site $\mathbf{s_{0}}$ with those at the
gauged site $\mathbf{s_{j}},$ $j\in\{1,\ldots,n\},$ respectively.
Note that both (\ref{pred:var:corr:prop1}) and
(\ref{pred:var:corr:prop2}) hold for $i=0.$ The properties of the
correlation structure in
(\ref{pred:var:corr:prop1})--(\ref{pred:var:corr:prop2}), lead us to
the conjecture that the model's predictive bands should increase
monotonically over time as more data become available, in the
absence of restrictions on $t\sigma_{\delta}^{2} = O(1)$ suggested
above. Furthermore, even conditioning on all the data, the
predictive bands should also increase over time. In support of these
conjectures, we prove that they hold in a simple case where $n=1$
and $T=2$ in (\ref{pred:var:obs:eqn})--(\ref{pred:var:state:eqn}).


\begin{theorem}\label{pred:var:simple:dlm:result}
For the first--order polynomial DLM in
(\ref{pred:var:obs:eqn})--(\ref{pred:var:state:eqn}) with $n=1$ and
$T=2$, assume the prior for $\beta_0$ to be $N(0, \sigma_\beta^2).$
The joint distribution of ${\mathbf{y}} = ( y_{01}, y_{11}, y_{02},
y_{12} )^\prime$ is $N({\mathbf{0}}, \Sigma),$ where
$$
\Sigma = (\sigma_{\beta}^{2} +
\sigma_{\delta}^{2}){\mathbf{1}_{4}}^{\prime}{\mathbf{1}_{4}} +
\mbox{ block--diagonal}\{ \sigma_{\varepsilon}^{2}\exp(-V/\lambda),
\sigma_{\delta}^{2}{\mathbf{1}_{2}}^{\prime}{\mathbf{1}_{2}} +
\sigma_{\varepsilon}^{2}\exp(-V/\lambda)\},
$$
${\mathbf{1}}_{k}^{\prime}$ being the $k \times 1$ vector of 1s
($k=1, 2, \ldots$). Then we have the following predictive
conditional variances:
\begin{eqnarray}
Var(y_{01}|y_{11}) & = & \frac{(\sigma_\beta^2 + \sigma_\delta^{2} +
\sigma_\varepsilon^2)^2 - (\sigma_\beta^2 + \sigma_\delta^2 +
\sigma_\varepsilon^2\exp(-d_{01}/\lambda))^{2}}
{\sigma_\beta^2 + \sigma_\delta^2 + \sigma_\varepsilon^2}; \label{pe:y01:y11}\\ 
\mbox{Var($y_{02}|y_{12}$)} & = & \frac{ (\sigma_{\beta}^{2} +
2\sigma_{\delta}^{2} + \sigma_{\varepsilon}^{2} )^{2} - (
\sigma_{\beta}^{2} + 2\sigma_{\delta}^{2} +
\sigma_{\varepsilon}^{2}\exp(-d_{01}/\lambda) )^{2}  } {
\sigma_{\beta}^{2} + 2\sigma_{\delta}^{2} + \sigma_{\varepsilon}^{2}
};
\label{pe:y02:y12} 
\end{eqnarray}
\begin{eqnarray}
Var(y_{01}|y_{11}, y_{12}) & = & \frac{M_{1}}{\Delta}; \label{pe:y01:y11:y12}\\ 
Var(y_{02}|y_{11}, y_{12}) & = & \frac{M_2}{\Delta}; \label{pe:y02:y11:y12} 
\end{eqnarray}
where
\begin{eqnarray}\label{pred:var:example:M1}
  \Delta & = & ( \sigma_{\beta}^{2} + \sigma_{\delta}^{2}
  + \sigma_{\varepsilon}^{2} )( \sigma_{\beta}^{2} + 2\sigma_{\delta}^{2}
  + \sigma_{\varepsilon}^{2} ) - ( \sigma_{\beta}^{2} + \sigma_{\delta}^{2} )^{2}, \\
  M_{1} & = & (\sigma_{\beta}^{2} + 2\sigma_{\delta}^{2}
  + \sigma_{\varepsilon}^{2} )\{ ( \sigma_{\beta}^{2} + \sigma_{\delta}^{2}
  + \sigma_{\varepsilon}^{2} )^{2} - ( \sigma_{\beta}^{2} + \sigma_{\delta}^{2}
  + \sigma_{\varepsilon}^{2}\exp(-d_{01}/\lambda) )^{2}\} \nonumber \\
        &   & - 2(\sigma_{\beta}^{2} + \sigma_{\delta}^{2})^{2}
        ( \sigma_{\varepsilon}^{2} -  \sigma_{\varepsilon}^{2}\exp(-d_{01}/\lambda) ),
\end{eqnarray}
and
\begin{eqnarray}\label{pred:var:example:M2}
  M_2 & = & (\sigma_\beta^2 + \sigma_\delta^2 + \sigma_\varepsilon^2)
  \{(\sigma_\beta^2 + 2\sigma_\delta^2 + \sigma_\varepsilon^2)^2
  - (\sigma_\beta^2 + 2\sigma_\delta^2 + \sigma_\varepsilon^2\exp(-d_{01}/\lambda))^2\}
  \nonumber\\
      &   & - 2(\sigma_\beta^2 + \sigma_\delta^2)^2
      (\sigma_\varepsilon^2 - \sigma_\varepsilon^2\exp(-d_{01}/\lambda)).
\end{eqnarray}
\end{theorem}

For this simple case, we would expect the predictive variance of
$y_{01}$ based on more data collected over time to be no greater
than that of $y_{01}$ based on less, that is, $$ Var(y_{01}|y_{11})
\geq Var(y_{01}|y_{11}, y_{12}) $$ and $$ Var(y_{02}|y_{12}) \geq
Var(y_{02}|y_{11}, y_{12}). $$ Moreover, we would expect that, based
on the same amount of data, the predictive variance of $y_{01}$
would be no greater than that of $y_{02},$ that is, $$
Var(y_{01}|y_{11}, y_{12}) \leq Var(y_{02}|y_{11}, y_{12}). $$

Dou et al. (2007) prove these conjectures and provide other
comparisons of these predictive variances.  We conclude that the
predictive variance function is a monotonic increasing function of
time $t$ based on the same set of data. It decreases when more data or
equivalently, more time is involved. Furthermore, the difference
between these predictive variances decreases as $t$ increases. It
increases with time even when conditioning on the same dataset.

\begin{theorem}\label{pe:timepoint:2:site:2}
For the first--order polynomial DLM in Theorem
\ref{pred:var:simple:dlm:result}, we have the following properties
of the predictive conditional variances:
  \begin{equation}\label{pe:diff:y01:time}
    \mbox{Var($y_{01}|y_{11}$)} - \mbox{Var($y_{01}|y_{11}, y_{12}$)}
    = \frac{ \sigma_{\varepsilon}^{4} ( \sigma_{\beta}^{2}
    + \sigma_{\delta}^{2} )^{2} (1-\exp(-d_{01}/\lambda))^{2} }
    {\Delta(\sigma_{\beta}^{2} + \sigma_{\delta}^{2} +\sigma_{\varepsilon}^{2}) }
    \geq 0;
    \end{equation}
    \begin{equation}\label{pe:diff:y02:time}
    \mbox{Var($y_{02}|y_{12}$)} - \mbox{Var($y_{02}|y_{11}, y_{12}$)}
    = \frac{ \sigma_{\varepsilon}^{4} ( \sigma_{\beta}^{2}
    + \sigma_{\delta}^{2} )^{2} (1-\exp(-d_{01}/\lambda)^{2}) }
    { \Delta( \sigma_{\beta}^{2} + 2\sigma_{\delta}^{2} + \sigma_{\varepsilon}^{2} ) }
    \geq 0;
    \end{equation}
    \begin{equation}\label{pe:diff:y02:y01:y11:y12:spatial}
    \mbox{Var($y_{02}|y_{11}, y_{12}$)} - \mbox{Var($y_{01}|y_{11}, y_{12}$)}
    =\frac{\sigma_{\varepsilon}^{4}\sigma_{\delta}^{2}(1-\exp(-d_{01}/\lambda))^{2}}{\Delta}
    \geq 0;
    \end{equation}
    \begin{equation}\label{pe:diff:y02:y01:spatial}
    \mbox{Var($y_{02}|y_{12}$)} - \mbox{Var($y_{01}|y_{11}$)}
    = \frac{ \sigma_{\varepsilon}^{4}\sigma_{\delta}^{2} ( 1 - \exp(-d_{01}/\lambda) )^{2} }
    { ( \sigma_{\beta}^{2} + \sigma_{\delta}^{2} + \sigma_{\varepsilon}^{2} )
    ( \sigma_{\beta}^{2} + 2\sigma_{\delta}^{2} + \sigma_{\varepsilon}^{2} ) }
    \geq 0;
    \end{equation}
    \begin{equation}\label{pe:diff:y02:y01:marg}
    Var(y_{01}|y_{11}) - Var(y_{01}|y_{11}, y_{12})
    \geq Var(y_{02}|y_{12}) - Var(y_{02}|y_{11}, y_{12});
    \end{equation}
    \begin{equation}\label{pe:diff:y02:y01:marg:cond}
    Var(y_{02}|y_{12}) - Var(y_{01}|y_{11})
    \leq Var(y_{02}|y_{11}, y_{12}) - Var(y_{01}|y_{11}, y_{12}).
    \end{equation}
\end{theorem}

As an immediate consequence of
(\ref{pe:diff:y02:y01:y11:y12:spatial}), the predictive variances
increase monotonically at successive time points conditional on all
the data. That leads to monotonically increasing coverage
probabilities at the ungauged sites, an interesting phenomenon
discussed in Section \ref{problems:in:DLM}. There we will also
discuss the lessons learned in this section in relation to our
empirical findings.

Next, we present a curious result about the properties of the above
predictive variances that may explain some of their key features.
This result concerns these predictive variances as functions of
$\lambda,$ $d_{01}$ or $\sigma_{\varepsilon}^{2}.$ Part of its proof
is included in Appendix \ref{pred:var:Result3}.

\begin{corollary}\label{pred:var:key:result}
The predictive conditional variances in
(\ref{pe:y01:y11})--(\ref{pred:var:example:M2}) increase as $d_{01}$
increases, or $\sigma_{\varepsilon}^{2}$ increases, or $\lambda$
decreases.
\end{corollary}
Thus, keeping two parameters fixed, these predictive conditional
variances are monotone functions of the remaining one. Therefore,
the DLM can paradoxically lead to larger predictive variances when
conditioning on more data. For example, in the case $n=1$ and $T=2,$
applying the DLM model with only the data at $T=2$ yields the
predictive variance $Var^{*}(y_{02}|y_{12})$, which is exactly the
same as $Var(y_{01}|y_{11})$ in (\ref{pe:y01:y11}). This predictive
variance is smaller than $Var(y_{02}|y_{11}, y_{12})$ in
(\ref{pe:y02:y11:y12}), which is based on more data, under certain
condition specified in the next corollary.

\begin{corollary}\label{pred:var:paradox}
For the first--order polynomial DLM in Theorem 1,
\begin{eqnarray}\label{pred:var:paradox:condition}
Var^{*}(y_{02}|y_{12}) < Var(y_{02}|y_{11}, y_{12}) & \mbox{if and
only if} & \sigma_{\varepsilon}^{2} > \sigma_{\beta}^{2} \left( 1 +
\frac{\sigma_{\beta}^{2}}{\sigma_{\delta}^{2}} \right).
\end{eqnarray}
\end{corollary}
The behavior suggested by Corollary \ref{pred:var:paradox} is
actually observed in our application (see Section
\ref{problems:in:DLM}).

\section{Implementation}\label{chapter2:algo:est}
This section very briefly describes how to implement our model using
the MCMC method, more specifically, the
forward--filtering--backward--sampling algorithm of Carter and Kohn
(1994). The details are given by Dou et al. (2007).

\subsection{Metropolis--within--Gibbs algorithm}\label{chapter2:algo:est:mcmc}
The joint distribution, $p(\lambda, \sigma^{2}, x_{1:T},
y_{1:T}^{m}, a_{1}, a_{2} | y_{1:T}^{o})$, is the object of
interest.  Here $y_{1:T}^{o} = ({\mathbf{y_{1}^{o}}}, \ldots,
{\mathbf{y_{T}^{o}}})$ represents the observation matrix at the $n$
gauged sites up to time $T.$ Moreover, $x_{1:T} = (x_{1}, \ldots,
x_{T}): (2n+1) \times T$ is the vector of state parameters at the
$n$ gauged sites until time $T.$ For simplicity, the values of
$\gamma$ are fixed but the problem of setting them will be addressed
below. Additional detail can be found in Appendix
\ref{appendix:section:2:3:1}.

Since that joint distribution does not have a closed form, direct
sampling methods fail, leading to the use of the Markov Chain Monte
Carlo (MCMC) method. A {\it{blocking MCMC}} scheme increases
iterative sampling efficiency, three blocks being chosen for reasons
given in Dou et al. (2007): $(\lambda, \sigma^{2}, x_{1:T}), $
$y_{1:T}^{m}$ and $(a_{1}, a_{2}).$ More precisely we can:
\begin{enumerate}
\item[(i)] sample from $p(x_{1:T}, \lambda, \sigma^{2}|a_{1}, a_{2}, y_{1:T})$
\item[(ii)] sample from $p(y_{1:T}^{m} | \lambda, \sigma^{2}, x_{1:T}, a_{1},
a_{2}, y_{1:T}^{o})$ and
\item[(ii)] sample from $p(a_{1}, a_{2}|x_{1:T}, \lambda, \sigma^{2}, y_{1:T}).$
\end{enumerate}

Since $p(\lambda, \sigma^{2}, x_{1:T}|a_{1}, a_{2}, y_{1:T})$ has no
closed form, the full conditional posterior distribution of
$x_{1:T}$ is obtained by Kalman filtering and smoothing, in other
words, by the FFBS algorithm. Assuming an inverse Gamma hyperprior
for $\sigma^{2},$ the conditional posterior distribution of
$\sigma^{2}$ given the range and phase parameters is also inverse
Gamma distributed with new shape and scale parameters. Note that
\begin{eqnarray}\label{joint:post:range:variance:state}
p(\lambda, \sigma^{2}, x_{1:T}|a_{1}, a_{2}, y_{1:T}) & = &
p(\lambda|a_{1}, a_{2}, y_{1:T})p(\sigma^{2}|\lambda, a_{1}, a_{2},
y_{1:T})  \nonumber \\                                & & \times
p(x_{1:T}|\lambda, \sigma^{2}, a_{1}, a_{2}, y_{1:T}),
\end{eqnarray}
indicating that we can sample iteratively from the three conditional
posterior distributions on the right--hand--side of
(\ref{joint:post:range:variance:state}) to obtain samples from
$p(\lambda, \sigma^{2}, x_{1:T}|a_{1}, a_{2}, y_{1:T}).$ However,
$p(\lambda|a_{1}, a_{2}, y_{1:T})$ has no closed form, leading us to
sample $\lambda$ by a {\it{Metropolis--Hasting}} chain within a
Gibbs sampling cycle, an algorithm as described in the next three
subsections.

\subsection*{Sampling from $p(\lambda, \sigma^{2}, x_{1:T}|a_{1}, a_{2}, y_{1:T})$} \label{ffbs:state:sample}

To sample $(\lambda, \sigma^{2}, x_{1:T})$ from $p(\lambda,
\sigma^{2}, x_{1:T}|a_{1}, a_{2}, y_{1:T}),$ we use the block MCMC
scheme. Because of (\ref{joint:post:range:variance:state}), we could
ideally iteratively sample $\lambda$ from $p(\lambda|a_{1}, a_{2},
y_{1:T}),$ $\sigma^{2}$ from $p(\sigma^{2}|\lambda, a_{1}, a_{2},
y_{1:T})$ and $x_{1:T}$ from $p(x_{1:T}|\lambda, \sigma^{2}, a_{1},
a_{2}, y_{1:T}).$ However, because we do not have a closed form for
the posterior density of $p(\lambda|a_{1}, a_{2}, y_{1:T})$, we use
instead the {\it{Metropolis--Hasting algorithm}} to sample
$\lambda$, given the data, from the following a quantity that is
proportional to its posterior density, that is,
\begin{eqnarray}
p(\lambda|a_{1}, a_{2}, y_{1:T}) & \propto &
p(\lambda)\prod_{t=1}^{T}|Q_{t}|^{-\frac{1}{2}}\left[ \beta +
\frac{1}{2} \sum_{t=1}^{T}
{\mathbf{e_{t}}}^{\prime}Q_{t}^{-1}{\mathbf{e_{t}}}
\right]^{-(nT/2+\alpha)}.
\end{eqnarray}
Since we cannot compute the normalization constant for
$p(\lambda|a_{1}, a_{2}, y_{1:T}),$ the Metropolis--Hasting
algorithm is used. The proposal density, $q(.,.),$ is selected to be
a lognormal distribution, because the parameter space is bounded
below by $0$, making the Gaussian distribution inappropriate. As
Moller (2003) points out, this alternative to a random walk
Metropolis considers the proposal move to be a random multiple of
the current state. From the current state $\lambda^{(j-1)} (j>1),$
the proposed move is $\lambda^{*} = \lambda^{(j-1)}e^{Z},$ where $Z$
is drawn from a symmetric density, such as normal. In other words,
at iteration $(j),$ we sample a new $\lambda^{*}$ from this proposal
distribution, centered at the previously sampled $\lambda^{(j-1)}$
with a tuning parameter, $\tau^{2}$, as the variance for the
distribution of $Z$. Gamerman (2006) suggests the acceptance rate,
that is, the ratio of accepted $\lambda^{*}$ to the total number of
iterations, be around $50\%.$ We tune $\tau^{2}$ to attain that
rate. If the acceptance rate were too high, for example, $70\%$ to
$100\%,$ we would increase $\tau^{2}.$ If too low, for example, $0$
to $20\%,$ we would decrease $\tau^{2}$, to narrow down the search
area for $\lambda^{*}.$

The Metropolis--Hasting algorithm proceeds as follows. Given
$\lambda^{(j-1)},$ $a_{1}^{(j-1)},$ $a_{2}^{(j-1)}$ and
$y_{1:T}^{(j-1)}$, where $j>1:$
\begin{enumerate}
\item[$\bullet$] Draw $\lambda^{*}$ from $LN(\lambda^{(j-1)}, \tau^{2}).$
\item[$\bullet$] Compute the acceptance probability: $$
\alpha(\lambda^{(j-1)},\lambda^{*}) = \min{\left\{1,
\frac{p(\lambda^{*}|a_{1}^{(j-1)}, a_{2}^{(j-1)}, y_{1:T}^{(j-1)})
/q(\lambda^{(j)}, \lambda^{*}) }{p(\lambda^{(j-1)}| a_{1}^{(j-1)},
a_{2}^{(j-1)}, y_{1:T}^{(j-1)}) /q(\lambda^{*}, \lambda^{(j-1)})}
\right\} }.$$
\item[$\bullet$] Accept $\lambda^{*}$ with probability
$\alpha(\lambda^{(j-1)}, \lambda^{*}).$ In other words, sample $u
\sim U[0,1]$ and let $\lambda^{(j)} = \lambda^{*}$ if
$\lambda^{*}<u$ and $\lambda^{(j)} = \lambda^{(j-1)}$ otherwise.
\end{enumerate}
We run this algorithm iteratively until convergence is reached.

Next, we sample $\sigma^{2}$  given the accepted $\lambda$'s,
$a_{1}$'s, $a_{2}$'s and $y_{1:T}$. The prior for $\sigma^{2}$ is
chosen to be an inverse gamma distribution with shape parameter
$\alpha$ and scale parameter $\beta.$ The posterior distribution for
$\sigma^{2}$ is also an inverse gamma distribution, but with a shape
parameter $\alpha + \frac{nT}{2}$ and a scale parameter $\beta +
\frac{1}{2}\sum_{t=1}^{T}{\mathbf{e_{t}}}^{\prime}
Q_{t}^{-1}{\mathbf{e_{t}}}.$

We now sample $x_{1:T}$ given the accepted $\lambda$'s,
$\sigma^{2}$'s, phase parameters and $y_{1:T},$ using FFBS. West and
Harrison (1997) propose a general theorem for inference about the
parameters in the DLM framework. For time series data, the usual
method for updating and predicting is the Kalman filter. Dou et al.
(2007) present a FFBS algorithm (similar to the Kalman filter
algorithm) to resample the state parameters conditional on all the
other parameters and observations as part of the MCMC method for
sampling $x_{1:T} = ({\mathbf{x_{1}}},\ldots, {\mathbf{x_{T}}})$
from the smoothing distribution \newline $p({\mathbf{x_{t}}} |
\lambda, \sigma^{2}, a_{1}, a_{2}, y_{1:T}).$

The initial state parameter is given by
\begin{equation}\label{dlm:init}
({\mathbf{x_{0}}} | {\mathbf{y_{0}}}, {\mathbf{\theta}}) \sim
N[{\mathbf{m_{0}}}, \sigma^{2}C_{0}],
\end{equation}
where ${\mathbf{y_{0}}}$ being the initial information, with
${\mathbf{m_{0}}}$ and $C_{0}$ known. Later in Section
\ref{chapter2:example:Cluster2}, we consider how to set them for
Cluster 2 AIRS dataset (1995). Let ${\mathbf{\theta}} = (\lambda,
\sigma^{2}, a_{1}, a_{2}, {\mathbf{\gamma}}).$  Now suppose for
expository purposes, that all the prior information has been given
and ${\mathbf{\theta}}$'s coordinates are mutually independent.

\subsection*{Sampling from $p(y_{1:T}^{\mbox{\tiny{m}}} |
\lambda, \sigma^{2}, x_{1:T}, a_{1}, a_{2},
y_{1:T}^{\mbox{\tiny{o}}})$} \label{ffbs:missing:sample}

MCMC can be used to fill in missing values at each iteration. To see
how, note that at any fixed time point $t$, after appropriately
defining a scale matrix $R_{t},$ we can rewrite the observation
vector ${\mathbf{y_{t}}}$ as follows: $$ R_{t}{\mathbf{y_{t}}} =
\left(
                           \begin{array}{c}
                               y_{t}^{m}\\
                               y_{t}^{o}
                           \end{array}
                         \right), $$
where $y_{t}^{m} : n_{t}\times 1$ denotes the missing response(s) at
time $t$ and $y_{t}^{o}: (n-n_{t})\times 1$ the observed response(s)
at $t.$ Notice that ``o'' represents ``observed'' and ``m'',
``missing''.

Let $R_{t} = ({\mathbf{e}}_{n_{1}}, \ldots, {\mathbf{e}}_{n_{t}},
{\mathbf{e}}_{k_{1}}, \ldots, {\mathbf{e}}_{k_{n-n_{t}}})^{\prime},$
where $\{ {\mathbf{s}}_{n_{j}}: j=1,\ldots,t\}$ represents the set
of gauged sites containing missing values at time point $t,$ $\{
{\mathbf{s}}_{k_{j}}: j=1,\ldots, n-n_{t} \}$ the set of gauged
sites containing observed values at time $t,$ for all
$t=1,\ldots,T;$ and ${\mathbf{e}}_{j} = (e_{j1}, \ldots,
e_{jn})^{\prime}: n \times 1$ such that $e_{jk}=I_{j=k}$ for
$k=1,\ldots,j$ and $j \in \mathcal{Z^{+}}.$

We already know that
\begin{eqnarray*}
( {\mathbf{y_{t}}} | \lambda, \sigma^{2}, {\mathbf{x_{t}}},
{\mathbf{a}} ) & \sim & N[ F_{t}^{\prime}{\mathbf{x_{t}}},
\sigma^{2}\exp\{ - V/\lambda\}],
\end{eqnarray*}
so that $R_{t}{\mathbf{y_{t}}}$ is also multivariate normally
distributed, that is,
$$
\begin{array}{lclcl}
(R_{t}{\mathbf{y_{t}}} | \lambda, \sigma^{2}, {\mathbf{x_{t}}},
{\mathbf{a}}) & = & ( ( {\mathbf{y_{t}^{m}}}, {\mathbf{y_{t}^{o}}}
)^{\prime} | \lambda, \sigma^{2}, {\mathbf{x_{t}}}, {\mathbf{a}} ) &
\sim & N[ {\mathbf{\tilde{\mu}_{t}}}, \tilde{\Sigma}_{t}],
\end{array}
$$
where
\begin{eqnarray*}
{\mathbf{\tilde{\mu}_{t}}} & = & R_{t}F_{t}^{\prime}{\mathbf{x_{t}}} \\
\tilde{\Sigma}_{t}         & = &
\sigma^{2}R_{t}\exp\{-V/\lambda\}R_{t}^{\prime}.
\end{eqnarray*}

We can also partition ${\mathbf{\tilde{\mu}_{t}}}$ as
${\mathbf{\tilde{\mu}_{t}}} = (
{\bf{\tilde{\mu}_{t}^{m}}}{}^{\prime},
{\bf{\tilde{\mu}_{t}^{o}}}{}^{\prime})^{\prime},$ where
${\mathbf{\tilde{\mu}^{m}_{t}}}: n_{t} \times 1$ and
${\mathbf{\tilde{\mu}^{o}_{t}}}: (n - n_{t}) \times 1.$ Similarly,
we have
\begin{eqnarray*}
\tilde{\Sigma}_{t} & = &
\left(
        \begin{array}{cc}
          \tilde{\Sigma}^{mm}_{t} & \tilde{\Sigma}^{mo}_{t}\\
          \tilde{\Sigma}^{om}_{t} & \tilde{\Sigma}^{oo}_{t}
        \end{array}
\right),
\end{eqnarray*}
where $\tilde{\Sigma}^{mm}_{t}: n_{t} \times n_{t},$
$\tilde{\Sigma}^{mo}_{t}: n_{t} \times (n-n_{t})$ and
$\tilde{\Sigma}^{oo}_{t}: (n-n_{t}) \times (n-n_{t}).$

By a standard property of the multivariate normal distribution, we have
\begin{eqnarray}\label{miss:mcmc:distribution}
({\mathbf{y_{t}^{m}}} | \lambda, \sigma^{2}, {\mathbf{x_{t}}},
{\mathbf{a}}, {\mathbf{y_{t}^{o}}} )& \sim & N[
{\mathbf{\mu^{**}_{t}}}, \Sigma^{**}_{t}],
\end{eqnarray}
where
\begin{equation}\label{miss:mcmc:mean}
{\mathbf{\mu^{**}}} = {\mathbf{\tilde{\mu}^{m}_{t}}} +
\tilde{\Sigma}^{mo}_{t} ( \tilde{\Sigma}^{oo}_{t} )^{-1} (
{\mathbf{y_{t}^{o}}} - {\mathbf{\tilde{\mu}_{t}^{o}}} ),
\end{equation}
and
\begin{equation}\label{miss:mcmc:Sigma}
\Sigma^{**}_{t} = \tilde{\Sigma}^{mm}_{t} -
\tilde{\Sigma}^{mo}_{t}(\tilde{\Sigma}^{oo}_{t} )^{-1}
\tilde{\Sigma}^{om}_{t},
\end{equation}
for $t=1,\ldots,T.$

At each iteration, we draw ${\mathbf{y_{t}^{m}}}$ from the
corresponding distribution (\ref{miss:mcmc:distribution}) at each
time point $t$ and then we can write the response variables as
$y_{1:T} = (y_{1:T}^{m}, y_{1:T}^{o})$
where $y_{1:T}^{m} = (y_{1}^{m}, \ldots, y_{T}^{m})$ and
$y_{1:T}^{o} = (y_{1}^{o}, \ldots, y_{T}^{o}).$

\subsection*{Sampling from $p(a_{1}, a_{2}|\lambda, \sigma^{2},
x_{1:T}, y_{1:T})$} \label{ffbs:phase:sample}

We now present our method for sampling the phase parameters
${\mathbf{a}} = (a_{1}, a_{2})^{\prime}$ from its full conditional
posterior distribution, that is, $p({\mathbf{a}} | \lambda,
\sigma^{2}, x_{1:T}, y_{1:T}),$ by using the samples of $\lambda$,
$\sigma^{2}$ and $x_{1:T}$. For simplicity, we use the notation for
models (\ref{dlm:y})--(\ref{dlm:alpha}) in this section.

We then sample the constant phase parameters conditional on all the
other parameters and observations. Suppose ${\mathbf{a}} = (a_{1},
a_{2})^{\prime}$ has a conjugate bivariate normal prior with mean
vector ${\mathbf{\mu}}^{o} = (\mu_{1o}, \mu_{2o})^{\prime}$ and
covariance matrix $\Sigma^{0}.$ Then the posterior conditional
distribution for $\mathbf{a}$ is normal with mean vector
${\mathbf{\mu}}^{*}$ and covariance matrix $\Sigma^{*},$ where
${\mathbf{\mu}}^{*}$ and $\Sigma^{*}$ can be obtained from equations
given in Dou et al. (2007).

We will not use a non--informative prior such as $p({\mathbf{a}})
\propto 1 $ for $\mathbf{a}$ since that choice can lead to
non--identified posterior means or posterior variances. In fact for
that choice we find the posterior conditional distribution of
$\mathbf{a}$ to be normal with mean vector ${\mathbf{\mu}}=(\mu_{1},
\mu_{2})^{\prime}$ and covariance matrix $\Sigma$ from equations
given in Dou et al. (2007) along with the elements of $\Sigma$,
where $\Sigma$ can be singular for any $t = 12k$ ($k \in
\mathbb{Z}$). Hence, we obtain the extreme values at times $12, 24,
\ldots, 2880,$ that invalidates the assumption of constant phase
parameters across all the time scales when we sample from its full
conditional posterior distribution.

For fixed values of  $\lambda,$ $\sigma^{2}$ and $x_{1:T},$ we
sample the model parameter $\mathbf{a}=(a_{1}, a_{2})$ from
$N({\mathbf{\mu}}^{*}, \Sigma^{*})$ at each time point, and then
obtain the ``sample'' of $\mathbf{a}$ at this iteration by the
median of these samples across all the time points, under the
assumption that $(a_{1}, a_{2})$ are constant phase parameters in
the models (\ref{dlm:obs})--(\ref{dlm:state}).

\subsection{Summary}\label{mcmc:algorithm:summary:table}

The MCMC algorithm we use here resembles that of Huerta et al.
(2004), one difference being that we unlike them, use all the
samples after the burn--in period, not just the chain containing the
accepted samples. We believe the Markov chains of only accepted
results will lead to biased samples, thereby changing the detailed
balance equation of the Metropolis--Hasting algorithm.

The above algorithm we use for Cluster 2 AIRS dataset is summarized
as follows:

\vspace*{+0.1cm}
\begin{enumerate}
\item[1.] Initialization: sample
  \begin{eqnarray*}
    \lambda^{(1)}      & \sim & IG(\alpha_{\lambda}, \beta_{\lambda})\\
    {\sigma^{2}}^{(1)} & \sim & IG(\alpha_{\sigma^{2}}, \beta_{\sigma^{2}})\\
    x_{1:T}^{(1)}      & \sim & N(m_{0}, {\sigma^{2}}^{(1)}C_{0}).
  \end{eqnarray*}

\item[2.] Given the $(j-1)^{\mbox{\tiny{th}}}$ value $\lambda^{(j-1)},$
${\sigma^{2}}^{(j-1)},$ $x_{1:T}^{(j-1)},$ ${ y_{1:T}^{m} }^{(j-1)},$
$a_{1}^{(j-1)},$ $a_{2}^{(j-1)}$ and the observations $y_{1:T}^{o}$:
  \begin{enumerate}
    \item[(1)] Sample $(\lambda^{(j)}, {\sigma^{2}}^{(j)}, x_{1:T}^{(j)})$
    from $p(\lambda, \sigma^{2}, x_{1:T}| a_{1}^{(j-1)}, a_{2}^{(j-1)},
    y_{1:T}^{(j-1)}),$ where $$y_{1:T}^{(j-1)} = ({y_{1:T}^{m}}^{(j-1)},
    y_{1:T}^{o}).$$

     \begin{enumerate}
      \item[(i)]
       \begin{enumerate}
          \item[$\bullet$] Generate a candidate value $\lambda^{*}$
          from a logarithm proposal distribution $q(\lambda^{(j-1)},
          \lambda),$ that is, $LN(\lambda^{(j-1)}, \tau^{2})$ for some
          suitable tuning parameter $\tau^{2}.$

          \item[$\bullet$] Compute the acceptance ratio $\alpha(\lambda^{(j-1)},
          \lambda^{*})$ where $$
             \alpha(\lambda^{(j-1)}, \lambda^{*}) = \min\left\{ 1,
             \frac{ p(\lambda^{*}|a_{1}^{(j-1)}, a_{2}^{(j-1)}, y_{1:T}^{(j-1)})
             \lambda^{*} }{ p(\lambda^{(j-1)} |a_{1}^{(j-1)}, a_{2}^{(j-1)},
             y_{1:T}^{(j-1)} ) \lambda^{(j-1)} } . \right\}$$

          \item[$\bullet$] With probability $\alpha(\lambda^{(j-1)},
          \lambda^{*})$ accept the candidate value and set $\lambda^{(j)} =
          \lambda^{*};$ otherwise reject and set $\lambda^{(j)} = \lambda^{(j-1)}.$
       \end{enumerate}

     \item[(ii)] Sample ${\sigma^{2}}^{(j)}$ from $p(\sigma^{2}|\lambda^{(j)},
     a_{1}^{(j-1)}, a_{2}^{(j-1)}, y_{1:T}^{(j-1)}).$

     \item[(iii)] Sample $x_{1:T}^{(j)}$ from $p(x_{1:T}|\lambda^{(j)},
     {\sigma^{2}}^{(j)}, a_{1}^{(j-1)}, a_{2}^{(j-1)}, y_{1:T}^{(j-1)} ).$
     \end{enumerate}
    \item[(2)] Sample ${y_{1:T}^{m}}^{(j)}$ from $p(y_{1:T}^{m} | \lambda^{(j)},
    {\sigma^{2}}^{(j)}, x_{1:T}^{(j)}, a_{1}^{(j-1)}, a_{2}^{(j-1)}, y_{1:T}^{o} ).$

    \item[(3)] Sample $(a_{1}^{(j)}, a_{2}^{(j)})$ from $p(a_{1}, a_{2} |
    \lambda^{(j)}, {\sigma^{2}}^{(j)}, x_{1:T}^{(j)}, y_{1:T}^{(j)}),$
    where $ y_{1:T}^{(j)} = ({y_{1:T}^{m}}^{(j)}, y_{1:T}^{o}).$
   \end{enumerate}
\item[3.] Repeat until convergence.
\end{enumerate}
We have developed software to implement the DLM approach of this
section. To enhance the Metropolis--within--Gibbs algorithm, we
augment the R code with C to speed up the computation. The current
version, {\it GDLM.1.0}, is freely available at
http://enviro.stat.ubc.ca for different platforms such as Windows,
Unix and Linux.

\section{Interpolation and prediction}\label{chapter2:algo:pred}

This section describes how to interpolate hourly ozone
concentrations at ungauged sites using the DLM and the simulated
Markov chains for the model parameters (see Section
\ref{chapter2:algo:est}). In other words, suppose ${\mathbf{s_{1}}},
\ldots, {\mathbf{s_{u}}}$ are $u$ ungauged sites of interest within
the geographical region of Cluster 2 sites (excluding the
possibility of extrapolation). The objective is to draw samples from
$$
p(y_{1:T}^{s} | \lambda, \sigma^{2}, x_{1:T}, a_{1}, a_{2},
y_{1:T}),
$$
where $y_{1:T}^{s} = ({\mathbf{y_{1}^{s}}}, \ldots,
{\mathbf{y_{T}^{s}}}): 1 \times T$ and $y_{t}^{s}$ denotes the
unobserved square--root of ozone concentrations at the ungauged site
$\mathbf{s}$ and time $t,$ for $t=1,\ldots,T$ and for ${\mathbf{s}}
\in \{{\mathbf{s_{1}}}, \ldots, {\mathbf{s_{u}}}\}.$ Let
$(\alpha_{1t}^{s}, \alpha_{2t}^{s})$ denote the unobserved state
parameters at site $\mathbf{s}$ and time t. The DLM is given by
\begin{eqnarray}\label{spatial:dlm}
{\mathbf{y_{t}}}^{\mbox{\tiny{new}}} & =
&{{\mathbf{1}}_{n+1}}^{\prime}\beta_{t} +
S_{1t}(a_{1}){\mathbf{\alpha_{1t}}}^{\mbox{\tiny{new}}} +
S_{2t}(a_{2}){\mathbf{\alpha_{2t}}}^{\mbox{\tiny{new}}} +
{\mathbf{\nu_{t}}}^{\mbox{\tiny{new}}},
\end{eqnarray}
where ${\mathbf{y_{t}}}^{\mbox{\tiny{new}}} = (y_{t}^{s},
{\mathbf{y_{t}}}^{\prime})^{\prime},$
${\mathbf{\alpha_{t}}}^{\mbox{\tiny{new}}} = (\alpha_{1t}^{s},
{\mathbf{\alpha_{1t}}}^{\prime}, \alpha_{2t}^{s},
{\mathbf{\alpha_{2t}}}^{\prime})^{\prime},$ and
${\mathbf{\nu_{t}}}^{\textrm{\tiny{new}}} \sim N(0,
\sigma^{2}\exp(-V^{\textrm{\tiny{new}}}/\lambda)).$

In the following two subsections, we illustrate how to sample the
unobserved state parameters $\{( \alpha_{1t}^{s}, \alpha_{2t}^{s}
):t=1,\ldots,T\}$ from the corresponding conditional posterior
distribution, and demonstrate the spatial interpolation at the
ungauged site $\mathbf{s}$.

\subsection*{Sampling the unobserved state parameters}\label{chapter2:algo:pred:phase}

We first sample $\alpha_{jt}^{s}$ given $\alpha_{j,t-1}^{s},$
$\mathbf{\alpha_{jt}}$ and $\mathbf{\alpha_{j,t-1}}, j=1,2.$ From
the state equation (\ref{dlm:state}) for
${\mathbf{\alpha_{jt}}}^{\mbox{\tiny{new}}},$ we know that the joint
density of $\alpha_{jt}^{s}$ and $\mathbf{\alpha_{jt}}$ follows a
normal distribution, with covariance matrix
$\sigma^{2}\tau_{j}^{2}\exp{(-V^{\mbox{\tiny{new}}}/\lambda_{j})},$
where $V^{\mbox{\tiny{new}}}$ denotes the distance matrix for the
unobserved station and the monitoring stations. The conditional
posterior distribution, $$ p(\alpha_{jt}^{s} | \alpha_{j,t-1}^{s},
\lambda, \sigma^{2}, \beta_{t}, {\mathbf{\alpha_{1t}}},
{\mathbf{\alpha_{2t}}}, a_{1}, a_{2}, y_{1:T}),
$$
is derived in Appendix \ref{appendix:spatial}.

\subsection*{Spatial interpolation at ungauged sites}\label{chapter2:algo:pred:o3}

We interpolate the square--root of ozone concentration at the
ungauged sites by conditioning on all the other parameters and
observations at the gauged sites. As above, $y_{t}^{s}$ and
$\mathbf{y_{t}}$ are jointly normally distributed as a consequence
of the observation equation. The predictive conditional distribution
for $y_{t}^{s},$ that is, $p(y_{t}^{s} |\alpha_{1t}^{s},
\alpha_{2t}^{s}, \lambda, \sigma^{2}, \beta_{t},
{\mathbf{\alpha_{1t}}}, {\mathbf{\alpha_{2t}}}, a_{1}, a_{2},
y_{1:T}),$ is given in Appendix \ref{appendix:spatial}.

\section{Application}\label{chapter2:example:Cluster2}
\begin{figure}
\centering \makebox{\includegraphics[width=3in,height=5in,
angle=-90]{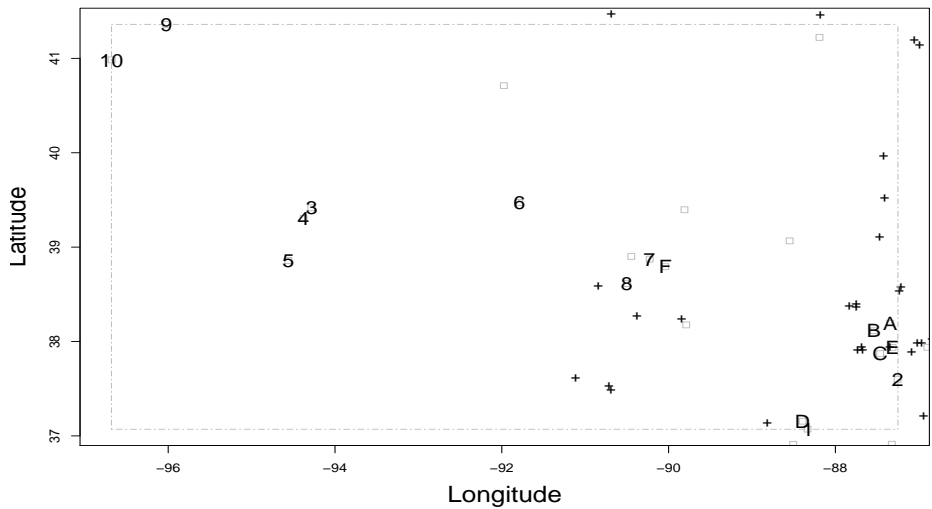}}
\caption{\label{fig:UG}Schematic representation of the locations of
ten gauged sites in Cluster 2 and the randomly chosen six ungauged
sites. (Number = Cluster 2 sites and letter = ungauged sites.)}
\end{figure}
This section applies our model to  the hourly ozone concentration
field described above. Six ungauged sites were randomly selected
from those available within the range of the sites in Cluster 2 to
play the role of ``unmonitored sites'' and help us assess the performance of
the DLM. The geographical locations of these six ungauged sites,
represented by the alphabetic letters, $A,\ldots, F,$ are shown in
Figure \ref{fig:UG}, along with the sites in Cluster 2.

\subsection{MCMC sampling}\label{chapter2:example:Cluster2:DLM}

This subsection presents a  MCMC simulation study in which samples
are drawn sequentially from the joint posterior distribution of the
model parameters in the DLM.

\vspace*{+0.5cm}
\noindent{\it{Initial settings}}\\
Following Huerta et al. (2004),  we use the following initial
settings for the starting values, hyperpriors and fixed model
parameters in the DLM:
\begin{enumerate}
\item[$\bullet$] The hyperprior for $\lambda$  is $IG(1, 5)$ and
for $\sigma^{2},$ $IG(2, 0.01).$ The expected value of $IG(1, 5)$
is $\infty$ and so are both of the variances of $p(\lambda)$ and
$p(\sigma^{2}).$ These vague priors for $\lambda$ and $\sigma^{2}$
are selected to reflect our lack of prior knowledge about their
distributions.
\item[$\bullet$] The initial information  for $x_{0},$ the initial
state parameter, is assumed to be normally distributed with mean vector
$\mathbf{m_{0}} = (2.85, -0.75\mathbf{1}_{n}^{\prime},
-0.08\mathbf{1}_{n}^{\prime})^{\prime}$ and covariance matrix
$\sigma_{1}^{2}C_{0},$ where $\sigma_{1}^{2} \sim IG(2, 0.01)$ and
$C_{0}$ is a block diagonal matrix with diagonal entries $1,$
$0.01\mathbf{1}_{n}^{\prime}$ and $0.01\mathbf{1}_{n}^{\prime}.$
\item[$\bullet$] The hyperprior for $\mathbf{a}$ is a bivariate
normal distribution with mean vector ${\mathbf{\mu^{o}}}=
(2.5, 9.8)^{\prime}$ and a diagonal matrix $\Sigma^{o}$ with
diagonal entries $0.5$ and $0.5.$
\item[$\bullet$] Some of the model parameters in the DLM are
fixed as follows: $\tau_{y}^{2} = 0.02,$ $\tau_{1}^{2} = 0.0002,$
$\tau_{2}^{2} = 0.0004,$ $\lambda_{1} = 25$ and $\lambda_{2} = 25.$
\end{enumerate}

\vspace*{+0.5cm}
\noindent{\it{Monitoring the convergence of the Markov chains}} \\

\begin{figure}
\centering
\makebox{\includegraphics[width=3in,height=5in,angle=-90]{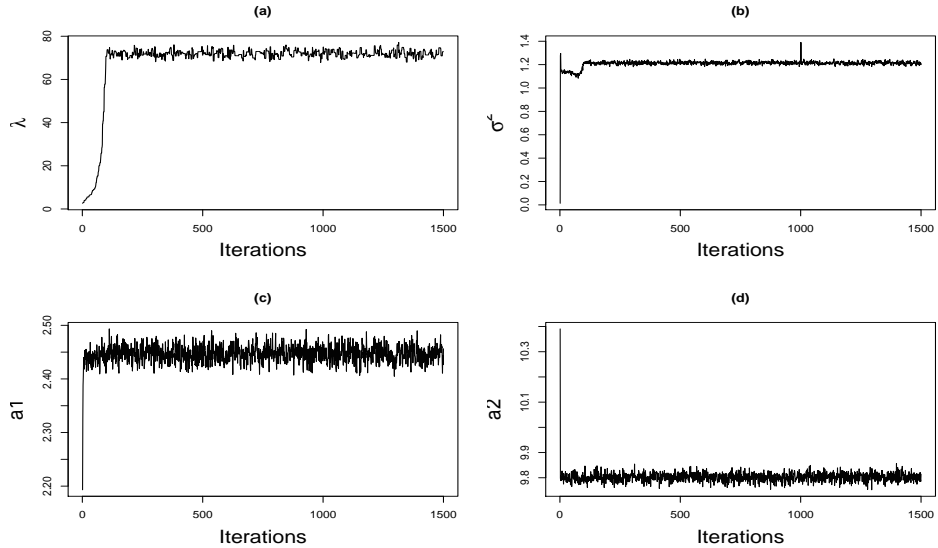}}
\caption{\label{MCMC:One:Chain:converge}Traces of model parameters
with the number of iterations of the Markov chains. The parameters
are: (a)--$\lambda,$ the range parameter; (b) --$\sigma^{2},$ the
variance parameter; (c) --$a_{1},$ the phase parameter with respect
to the $24$--hour periodicity; and (d) --$a_{2},$ the phase
parameter with respect to the $12$--hour periodicity.}
\end{figure}

Figure \ref{MCMC:One:Chain:converge} shows the trace plots of model
parameters $\lambda,$ $\sigma^{2},$ $a_{1}$ and $a_{2}$ with the
number of iterations of the simulated Markov chains where the total
number of iterations is $4,268.$ The burn--in period is chosen to be
$2,269$ and all the remaining Markov samples are collected for
posterior inference. The acceptance rate is approximately $62\%.$ We
observe that the Markov Chain converges after a run of less than
five hundreds iterations.

\begin{table}
\caption{\label{post:summary}Posterior
summaries for $\lambda,$ $\sigma^{2},$ $a_{1}$ and
$a_{2}.$}
\centering
\fbox{%
\begin{tabular}{llllll} \hline\hline
\mbox{Quantile} & $\lambda$ & $\sigma^{2}$ & $a_{1}$ & $a_{2}$  \\ \hline
\mbox{$2.5\%$}  & 69.29     & 1.19         & 2.42    & 9.77 \\
\mbox{Median}   & 71.83     & 1.21         & 2.45    & 9.80  \\
\mbox{$97.5\%$} & 75.37     & 1.24         & 2.48    & 9.84 \\ \hline
\end{tabular}}
\end{table}

Table \ref{post:summary} displays the median and $95 \%$ quantile
from the simulated Markov chains for the model parameters $\lambda,$
$\sigma^{2},$ $a_{1}$ and $a_{2}.$

\subsection{Spatial interpolation}\label{chapter2:example:Cluster2:Result}

This subsection assesses the model's performance by comparing the
interpolated values at the ungauged sites, $A,\ldots,F$, with the
measurements made there. We use the entire dataset to assess the
performance of the interpolation results. Table \ref{cov:prob} shows
the coverage probabilities of the credibility intervals (or
``credible intervals'' for short) for these six ungauged sites at
various norminal levels. Generally, the coverage probabilities at
the ungauged sites exceed their nominal levels indicating that the
error bands are too wide.

Among these six ungauged sites, Site $D$ has the highest coverage
probability seen in Table \ref{cov:prob}. This may be because of
$D$'s nearness to a close ``relative'' among the gauged sites,
namely, Site $1.$ That would be consistent with our assumption that
the spatial correlation is inversely proportional to the intersite
distance. At the same time, these unsatisfactory large coverage
probabilities point to a deficiency of the DLM.

\begin{table}
\caption{\label{cov:prob} Comparisons
between the empirical credible probability and the nominal levels
at the ungauged sites $A, \ldots,F.$}
\centering
\fbox{%
\begin{tabular}{cll} \hline\hline
\mbox{Nominal Prob (\%)} &
$\begin{array}{c}
  \mbox{Coverage Prob.s (\%)}\\
    \begin{array}{cccccc}
    A\hspace*{+0.4cm} & B\hspace*{+0.4cm} & C\hspace*{+0.4cm} & D\hspace*{+0.4cm} & E\hspace*{+0.4cm} & F\hspace*{+0.4cm}
    \end{array}
  \end{array}
 $  \\ \cline{1-2} \hline
\mbox{$95$} & $\begin{array}{cccccc}
94.9 & 96.9  & 96.5  & 99.7 & 96.1 & 98.1
\end{array}$
\\
\mbox{$90$} & $\begin{array}{cccccc}
91.9 & 93.7  & 93.5  & 99.4 & 93.6 & 96.8
\end{array}$
\\
\mbox{$80$} & $\begin{array}{cccccc}
84.8 & 88.5  & 88.2  & 97.7 & 89.6 & 94.3
\end{array}$
\\
\mbox{$70$} & $\begin{array}{cccccc}
                78.7 & 83.5  & 83.3  & 94.0 & 85.8 & 90.6
              \end{array}$ \\
\mbox{$60$} & $\begin{array}{cccccc}
                73.0 & 78.5  & 77.1  & 89.7 & 81.6 & 86.6
              \end{array}$ \\
\mbox{$50$} & $\begin{array}{cccccc}
                65.2 & 71.5  & 70.4  & 85.6 & 76.1 & 81.4
              \end{array}$ \\
\mbox{$40$} & $\begin{array}{cccccc}
               55.2 & 61.4  & 61.0  & 79.2 & 67.9 & 74.7
              \end{array}$ \\
\mbox{$30$} & $\begin{array}{cccccc}
               42.2 & 47.6  & 47.5  & 69.6 & 54.9 & 64.4
              \end{array}$ \\ \hline
\end{tabular}}
\end{table}

\begin{figure}
\centering \makebox{\includegraphics[width=3in,height=5in,angle=-90]
{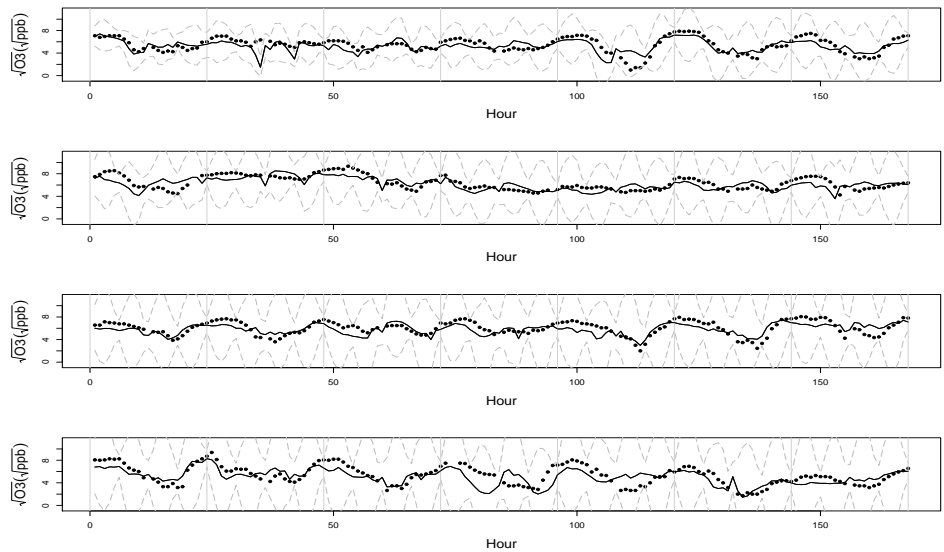}}
\caption{\label{UG:D:Wk1:Wk4}Interpolation at Ungauged Site D for
four successive weeks beginning from May 14, 1995. The square--root
of hourly ozone concentrations are plotted on the vertical axes,
hours on the horizontal axes.  The solid lines represent the
predicted median of the responses, the dashed lines represent the
$95 \%$ predictive intervals for the predicted square--root of ozone
concentrations and the solid dots represent the observations at
Ungauged Site $D$.}
\end{figure}

\begin{figure}
\centering
\makebox{\includegraphics[width=3in,height=5in,angle=-90]{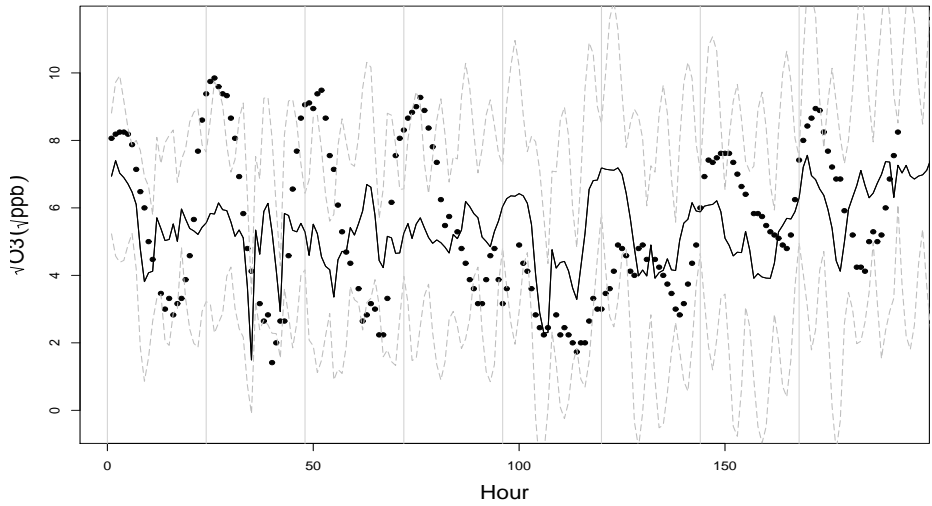}}
\caption{\label{UG:D:Wk17:Day120}Interpolation at Ungauged Site D
from the 17th week to the 120th day. The square--root of hourly
ozone concentrations are plotted in the vertical axes, hours on the
horizontal axes.}
\end{figure}

To explore this issue further, we compared the values predicted for
Ungauged Site $D$ from May 14 to September 11, 1995 and the
measurements made there. Figures \ref{UG:D:Wk1:Wk4} and in more
detail \ref{UG:D:Wk17:Day120}, which exemplify results reported in
more detail by Dou et al. (2007), depict the results for the first
four weeks and the last week of that period, respectively.
\begin{table}
\caption{\label{ungauge:gauge:friends}Close
``relatives'' of the ungauged sites, their global circle distance
(km) and the average of their correlation with their associated
gauged sites.}
\centering
\fbox{%
\begin{tabular}{cccc} \hline\hline
\mbox{Ungauged Site} &  \mbox{Relative(s)} & \mbox{GCD (km)} &
\mbox{Pearson's r} \\ \cline{1-4} \hline
$A$ & $2$     & $66.6$         & $0.73$ \\
$B$ & $2$     & $62.5$         & $0.74$ \\
$C$ & $2$     & $35.5$         & $0.84$ \\
$D$ & $1$     & $11.0$         & $0.95$ \\
$E$ & $2$     & $38.0$         & $0.70$ \\
$F$ & $(7,8)$ & $(18.6, 44.9)$ & $(0.84, 0.82)$ \\ \hline
\end{tabular}}
\end{table}

\begin{figure}
\centering
\makebox{\includegraphics[width=3in,height=5in,angle=-90]{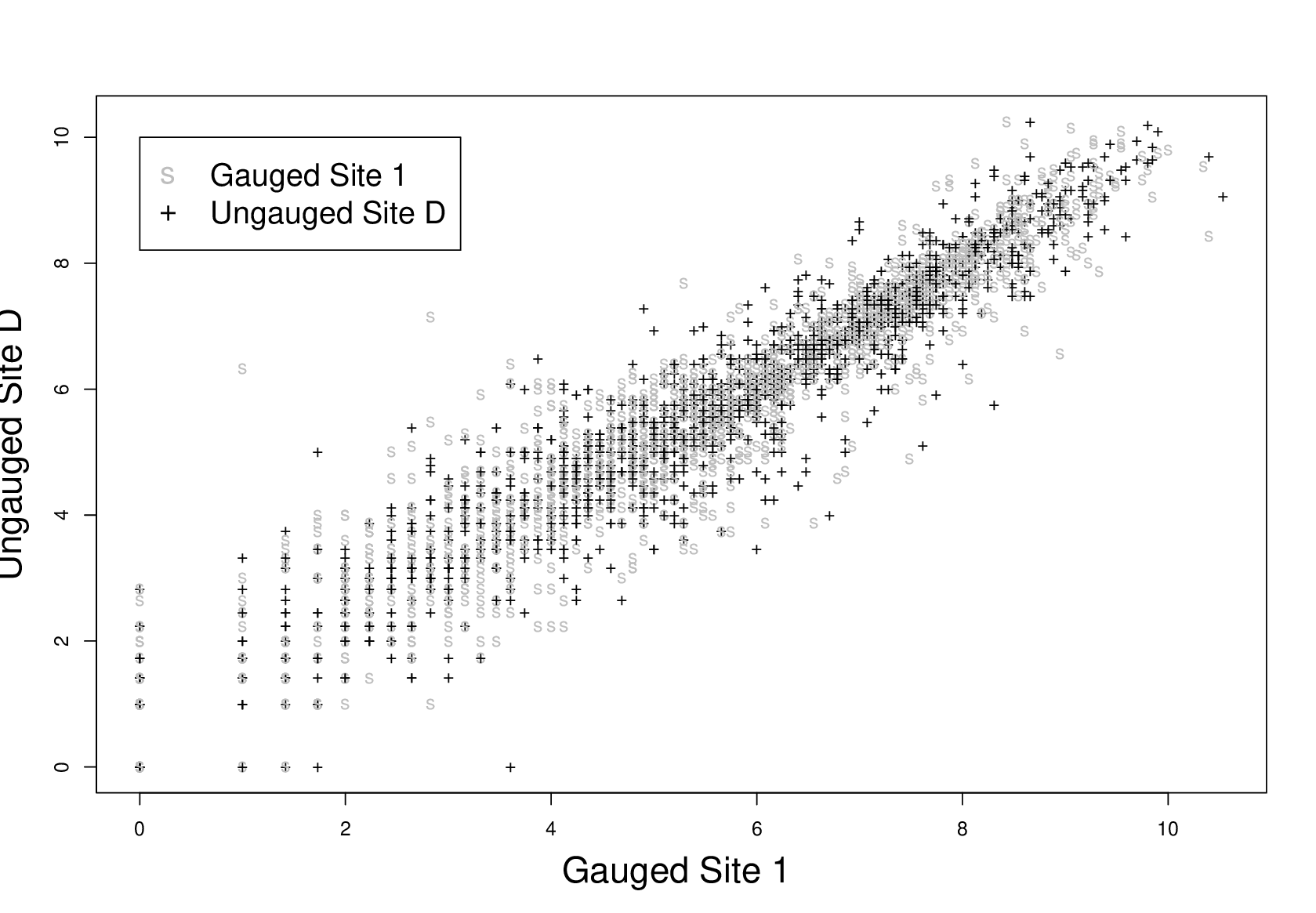}}
\caption{\label{Friend:UG:D:G:1}Scatterplot for the square--root of
ozone concentrations at Ungauged Site $D$ and its close relative,
Gauged Site $1.$ The square--root of hourly ozone concentrations are
plotted in both vertical and horizontal axes.}
\end{figure}

Furthermore, Table \ref{ungauge:gauge:friends} shows for all the
ungauged  sites, the close relatives they have among the Cluster 2
sites that lie within a radius of 100 km, the corresponding global
circle distance (GCD) in km, and along with the average of their
correlations. This table confirms that indeed $D$ does enjoy the
highest correlation with its relative. That relationship is further
explored in Figure \ref{Friend:UG:D:G:1} where we see a strong
linear relationship between Sites $D$ and $1$ as our coverage
probability assessment had suggested.

In spite of its reliance on the relatives, the DLM does not predict
responses at the ungauged sites very accurately as illustrated in
Figure \ref{UG:D:Wk17:Day120}. That points to problems with this
model which will be discussed in the next section.


\section{Discussion}\label{problems:in:DLM}

In general, the DLM provides a remarkably powerful modelling tool,
made practical by advances in statistical computing. However, its
substantial computational requirements still limits its
applicability. Moreover, the very flexibility that makes it so
powerful also imposes an immense burden of choice on the model. This
section summarizes critical issues and includes some suggestions for
improvement.

\vspace*{+0.8cm}
\noindent{\it{\textbf{Monitoring MCMC convergence}}}\\

\begin{figure}
\centering
\makebox{\includegraphics[width=3in,height=5in,angle=-90]{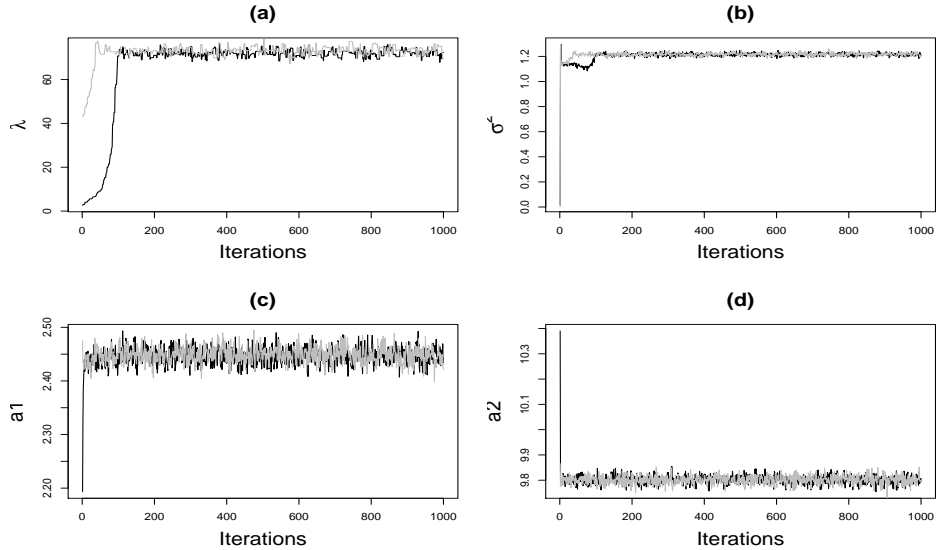}}
\caption{\label{MCMC:two:Chains}Traces of model parameters for a
number of iterations of two chains. The parameters are: (a)
--$\lambda,$ the range parameter; (b) --$\sigma^{2},$ the variance
parameter; (c)--$a_{1},$ the phase parameter with respect to the
$24$--hour periodicity; and (d) --$a_{2},$ the phase parameter with
respect to the $12$--hour periodicity.}
\end{figure}

Figure \ref{MCMC:two:Chains} represents the trace  plots of model
parameters $\lambda,$ $\sigma^{2},$ $a_{1}$ and $a_{2}$ of two
chains from the initial settings in Section
\ref{chapter2:example:Cluster2:DLM}. These two chains seem to mix
well after several hundreds iterations, suggesting at first glance
the Markov chains have converged.

\newpage
$\newline$ $\newline$
$\newline$\noindent{\it{\textbf{Autocorrelation and partial
autocorrelation of the simulated Markov chains}}}\\

However, we know that the autocorrelation, as measured by the
autocorrelation function (ACF), is very important when considering
the length of the chain. A highly auto--correlated chain needs a
long run to yield accurate estimates. Moreover, the partial
autocorrelation function (PACF) is also an important index for
assessing a Markov chain since large values of the PACF at lag $h$
indicates that the next value in the chain is dependent on past
values, not just on the most recent ones. \setcounter{figure}{6}
\begin{figure}
\centering
\makebox{\includegraphics[width=4in,height=5in,angle=-90]{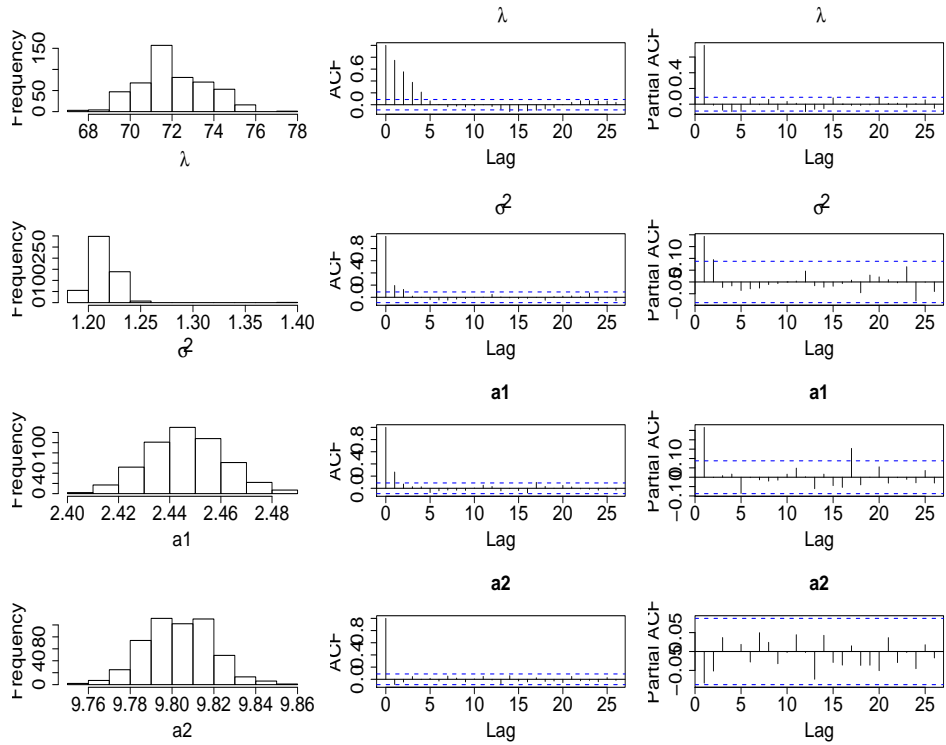}}
\caption{\label{MCMC:hist:acf:pacf}Histogram (left panel), ACF
(middle panel) and PACF (right panel) of model parameters of the
Markov chains after a burn--in period of $1,000.$ The parameters
are: (i) first row: --$\lambda,$ the range parameter; (ii) second
row: --$\sigma^{2},$ the variance parameter; (iii) third row:
--$a_{1},$ the phase parameter with respect to the $24$--hour
periodicity; and (iv) last row: --$a_{2},$ the phase parameter with
respect to the $12$--hour periodicity.}
\end{figure}

Figure \ref{MCMC:hist:acf:pacf} shows the histogram, ACF and PACF
plots for the Markov chains used in Section
\ref{chapter2:example:Cluster2:Result}, after a burn--in period of
$1,000.$ The ACF plots show the $\lambda$s to be highly
autocorrelated, in other words that the $\lambda$--chain does not
mix well, potentially leading to biased estimates in Section
\ref{chapter2:example:Cluster2:Result}. Thinning the chain might
reduce that autocorrelation. In other words, using every
$k^{\mbox{\tiny{th}}}$ ($k>1, k \in \mathcal Z^{+}$) $\lambda$
generated by the chain could be used to produce the estimates.
However, computational challenges make that strategy impractical; we
need to use the entire chain.

\vspace*{+0.8cm}
\noindent{\it{\textbf{Relationship between pairs of
$\lambda,$ $\sigma^{2},$ $a_{1}$ and $a_{2}$}}}\\

\begin{figure}
\centering
\makebox{\includegraphics[width=3in,height=5in,angle=-90]{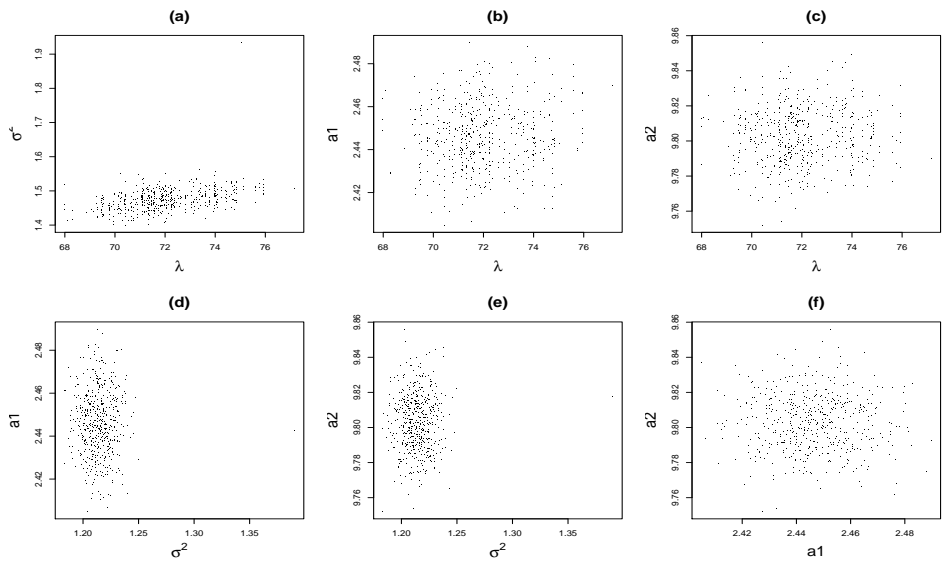}}
\caption{\label{scatter:model:parameters}Scatterplots for the pairs
of model parameters: (a) $\lambda$ v.s. $\sigma^{2};$ (b) $\lambda$
v.s. $a_{1};$ (c) $\lambda$ v.s. $a_{2};$ (d) $\sigma^{2}$ v.s.
$a_{1};$ (e) $\sigma^{2}$ v.s. $a_{2};$ and (f) $a_{1}$ v.s.
$a_{2}.$}
\end{figure}

Our prior assumptions make the model parameters $\lambda,$
$\sigma^{2},$ $a_{1}$ and $a_{2}$ uncorrelated. Figure
\ref{scatter:model:parameters} shows the relationship between the
pairs of these parameters as a way of investigating that assumption.
It seems valid except for the $\lambda$--$\sigma^{2}$ pair in graph
$(a).$ That graph shows a weak linear association between $\lambda$
and $\sigma^{2},$ thus pointing to a failure of that assumption for
that pair. Since $\sigma^2$ determines spatial variability while
$\lambda$ determines correlation this relationship seems intriguing.
Larger values of $\sigma^{2}$ tend to go with larger $\lambda$s,
i.e., diminished spatial correlation. Why they are coupled in this
way is unknown but it should be accounted for in future applications
of this model.

\vspace*{+.3cm}
\noindent{\it{\textbf{Time varying $\lambda$s and $\sigma^{2}$s:
empirical coverage probabilities versus nominal credible probabilities}}}\\
\begin{figure}
\centering \makebox{\includegraphics[width=3in,height=5in,
angle=-90]{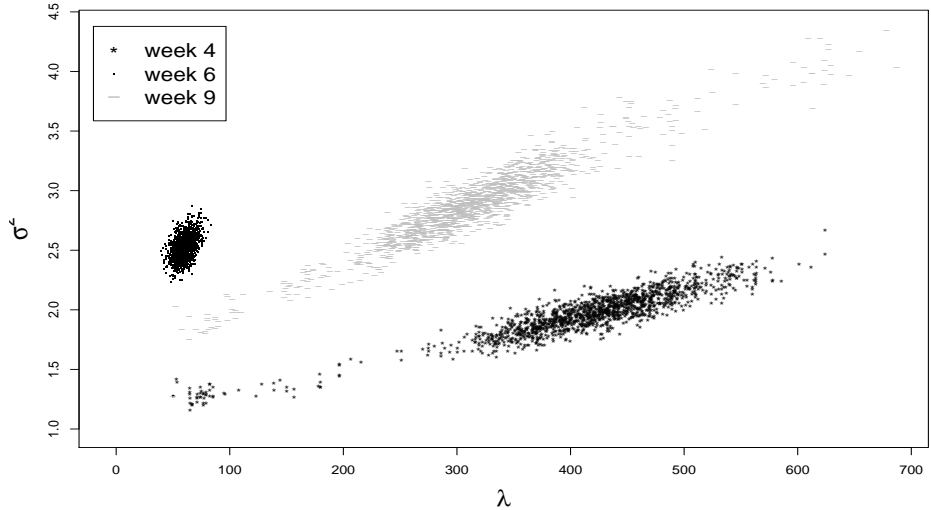}}
\caption{\label{time:vary:lambda:sigma}Scatterplots for $\lambda$
against $\sigma^{2}$ for various weeks, based on the MCMC samples
using one week's data, that is, weeks 4, 6 and 9, but starting from
the same initial values as those in Section
\ref{chapter2:example:Cluster2:DLM}.}
\end{figure}

Although we follow Huerta et al. (2004) in assuming the temporal
constancy of $\lambda$ and $\sigma^{2},$ it is natural to ask if
those generated by the MCMC method change over time. A variant of
this issue concerns the time domain of the application. Would the
results for these parameters change if we switched from one time
span to a longer one containing it? A ``yes'' to this question would
pose a challenge to anyone intending to apply the model, knowing
that the choice would have implications for the size of $\sigma^2$
and $\lambda$.

To address these concerns we carried out the  following studies:
\begin{enumerate}
\item[(i)] Study $\tilde{A}:$ Implement the DLM at ungauged sites
using weekly data ($W_{k}: k=1,\ldots,17$). Generate Markov chains
for $\lambda,$ $\sigma^{2},$ $a_{1}$ and $a_{2}.$ Obtain the coverage
probabilities at each ungauged site and week for fixed credibility
interval probabilities.

\item[(ii)] Study $\tilde{B}:$ Implement the DLM at ungauged sites
using week $1$ to week $17$ data ($W_{1:17} = \{W_{1}, \ldots,
W_{17}\}$). Estimate model parameters and interpolate the results
at those ungauged sites. Obtain the coverage probabilities at each
ungauged site and week for fixed credibility interval probabilities
using each week's data.

\item[(iii)] Study $\tilde{C}:$ Fix $\lambda_{k}^{*}$ at week
$k$ ($k=1,\ldots,17$) using values suggested by the Markov chains
generated in Study $\tilde{A}$. Then use these ${\mathbf{\lambda^{*}}}
= \{\lambda_{1}^{*}, \ldots, \lambda_{17}^{*}\}$ as fixed values
in the DLM to reduce computation time. In other words, go through
all the steps in the algorithm of Section
\ref{mcmc:algorithm:summary:table} but now using only fixed
$\lambda^{*}$s instead of generating them by a Metropolis--Hasting
step. (Note that we are then only using Gibbs sampling and an
MCMC blocking scheme.) Compute the corresponding coverage
probabilities using $W_{1:17}$ at each ungauged site and week for
fixed credibility interval probabilities.
\end{enumerate}

Studies $\tilde{A}$ and $\tilde{B}$ are intended  to explore the
effect of data and time propagation on the interpolation results.
Study $\tilde{C}$ aims to pick out any significant difference in the
interpolation results when using the fixed $\mathbf\lambda^{*}$
rather than using the Markov samples of $\lambda$s. It is also aimed
at finding how much time would be saved by avoiding the inefficient
Metropolis step. Table \ref{fix:lambda:weekdata} shows these fixed
$\mathbf\lambda^{*}$s used in Study $\tilde{C}.$ Table 5 shows the
time saved using fixed $\lambda^{*}$s against the one using the
Metropolis--Hastings algorithm.

\begin{table}
\caption{\label{fix:lambda:weekdata}Fixed values of $\mathbf\lambda^{*}$ in Study $\tilde{C}.$}
\setlength{\tabcolsep}{3.5pt}
\centering
\fbox{%
\begin{tabular}{|l|lllllllll|} \hline\hline
\mbox{\textbf{Week}}   & $1$   & $2$   & $3$   & $4$   & $5$   & $6$
& $7$   & $8$   & $9$   \\ \hline $\mathbf{\lambda^{*}}$ & 54.2  &
178.5 & 83.7  & 405.4 & 86.6  & 59.7 & 199.3 & 144.1 & 322.7 \\
\hline \hline \mbox{\textbf{Week}}   & $10$  & $11$  & $12$  & $13$
& $14$  & $15$ & $16$  & $17$  &       \\ \hline
$\mathbf{\lambda^{*}}$ & 142.2 & 172.7 & 187.9 & 315.8 & 419.0 &
99.8 & 260.3 & 284.8 &       \\ \hline
\end{tabular}}
\end{table}

\begin{table}
\caption{\label{time:saving}Summary for the computational time in
Studies $\tilde{A},$ $\tilde{B}$ and $\tilde{C}$. Time is measured
in seconds. The total is for a complete summer long MCMC run without
spatial prediction.} \centering
\fbox{%
\begin{tabular}{|cccc|cc|}\hline\hline
\multicolumn{4}{|c|}{}&\multicolumn{2}{|c|}{\textbf{Time (seconds)}}\\\hline
{\bf Study} & \textbf{Data} & \textbf{Iteration total} &
\textbf{Accept(\%)} & \textbf{Total} & \textbf{/Iteration} \\\hline
$\tilde{A}$ & $W_{k}$   & 1,500 & 0.82 & 17018  & 13.8  \\
$\tilde{B}$ & $W_{1:17}$ & 1,000 & 0.35 & 326782 & 932.3 \\
$\tilde{C}$ & $W_{1:17}$ & 1,000 & 1.00 & 329349 & 329.3 \\ \hline
\end{tabular}}
\end{table}

Figure \ref{time:vary:lambda:sigma} illustrates the MCMC estimation
results obtained in Study $\tilde{A}.$ It plots the Markov chains of
$\lambda$ and $\sigma^{2}$ using weekly data. It is obvious that
$\lambda$ and $\sigma^{2}$ vary from week to week, which implies
that the constant $\lambda$--$\sigma^{2}$ model is not tenable over
a whole summer for this dataset.

\setcounter{figure}{9}
\begin{figure}
\centering
\makebox{\includegraphics[width=3in,height=5in,angle=-90]{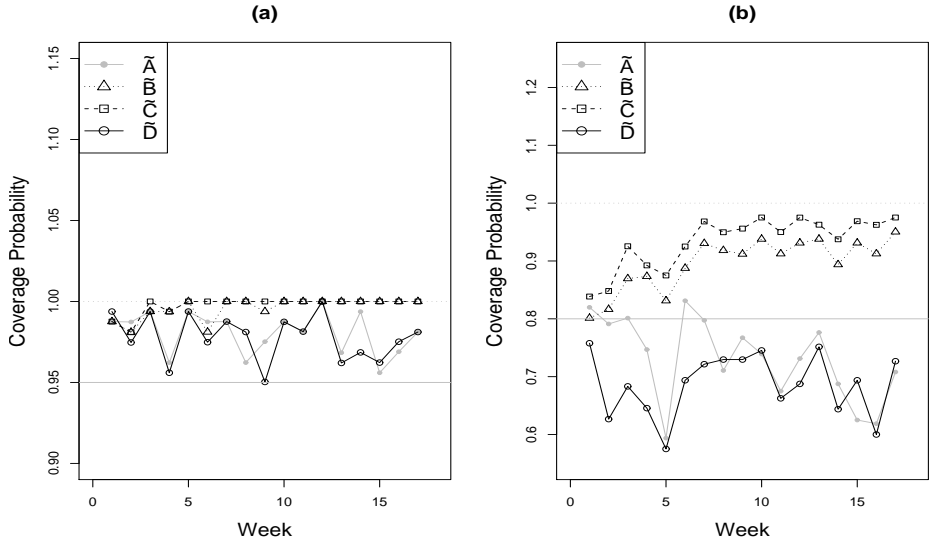}}
\caption{\label{cov:prob:paper}Coverage probability versus: (a)
$95\%$ nominal level for Ungauged Site D, and (b) $80\%$ nominal
level for Ungauged Site C. These coverage probabilities are computed
for {\it{Study $\tilde{A}$}}: weekly data (solid bullet with solid
line); {\it{Study $\tilde{B}$}}: $W_{1:17}$ (up-triangle with dotted
line); {\it{Study $\tilde{C}$}}: $W_{1:17}$ but with fixed
$\mathbf{\lambda^{*}}$ (square with dashed line); and {\it{Study
$\tilde{D}$}}: $W_{1:17}$ but with fixed $\mathbf{\lambda^{*}}$ and
modified $\tau_{y}^{2},$ $\tau_{1}^{2},$ and $\tau_{2}^{2}$ (empty
circle with solid line).}
\end{figure}

Figure \ref{cov:prob:paper} typifies figures in Dou et al. (2007)
showing the coverage probabilities for various predictive intervals
associated with the interpolators in these three studies. The solid
line with bullets represents the results for Study $\tilde{A},$ the
dotted line with up-triangles for Study $\tilde{B},$ and the dashed
line with squares for Study $\tilde{C}.$ These graphs show that the
coverage probabilities of Study ${\tilde B}$ are similar to that of
Study ${\tilde C}.$ This suggests that we could use the entries in
Table \ref{fix:lambda:weekdata} as fixed $\lambda^{*}$s in the DLM
to obtain interpolation results similar to those obtained using the
Metropolis--within--Gibbs algorithm.

We have studied the prediction accuracy of the simplest DLM, namely,
the first--order polynomial model, in Section
\ref{pred:var:simplest:dlm}. As a result, the predictive variances
should increase monotonically at successive time points conditional
on all the 17 weeks' data, in the general DLM setting (see Section
\ref{pred:var:simplest:dlm}). The plots exhibit a monotonic
increasing trend in the coverage probabilities of both Studies
$\tilde{B}$ and $\tilde{C}.$ This trend agrees with the graph of the
coverage probabilities in Figure \ref{cov:prob:paper}. Nevertheless,
those coverage probabilities of both studies deviate slightly from
the expected monotonically increasing trend at some time points
because of the time varying effect of $\lambda$--$\sigma^{2}$
monitored in Figure \ref{time:vary:lambda:sigma}.

On the other hand, Study $\tilde{C}$ enjoys significant
computational time savings compared with $\tilde{B}.$ Table
\ref{time:saving} suggests that the computation time of the former
is almost 2.8 times faster than the latter.

Study $\tilde{B}$ shows an intuitively unappealing increase in the
uncertainty of interpolation results as time increases; coverage
probabilities get larger over time as we see in Table
\ref{cp:threestudies:80cp}. This increase may be interpreted as
saying that for the DLM models, the $\lambda$s and $\sigma^{2}$s
collected from the data should vary over the entire time span of the
study, while the prior postulates that they do not vary over that
time span. The observed phenomenon may also be due to
mis--specification of the model parameter values $\gamma =
(\tau_{y}^{2}, \tau_{1}^{2}, \ldots, \lambda_{2})$ (See the initial
settings for $\gamma$ in Section
\ref{chapter2:example:Cluster2:DLM}.).

\begin{table}

\caption{\label{cp:threestudies:80cp}Coverage probabilities $(\%)$
for studies $\tilde{A},$ $\tilde{B}$ and $\tilde{C}$ at Ungauged
Sites A, B, and C, at $80\%$ nominal level.} \centering
\fbox{%
\begin{tabular}{|cccccccccc|} \hline\hline
\mbox{Ungauged Site} &  &A& &  &B& &  &C&
\\
\mbox{Study} & $\tilde{A}$ & $\tilde{B}$ & $\tilde{C}$ & $\tilde{A}$
& $\tilde{B}$ & $\tilde{C}$ & $\tilde{A}$ & $\tilde{B}$ &
$\tilde{C}$ \\ \hline
\mbox{Week 1} & 66 & 65 & 72 & 80 & 78 & 89 & 82 & 80 & 84  \\
\mbox{Week 2} & 73 & 71 & 80 & 76 & 78 & 83 & 79 & 81 & 85  \\
\mbox{Week 3} & 63 & 73 & 82 & 82 & 86 & 91 & 80 & 87 & 93  \\
\mbox{Week 4} & 57 & 74 & 81 & 66 & 83 & 88 & 75 & 87 & 89  \\
\mbox{Week 5} & 53 & 70 & 82 & 68 & 83 & 90 & 59 & 83 & 88  \\
\mbox{Week 6} & 73 & 80 & 88 & 75 & 83 & 89 & 83 & 89 & 93  \\
\mbox{Week 7} & 69 & 88 & 90 & 80 & 92 & 94 & 80 & 93 & 97  \\
\mbox{Week 8} & 66 & 89 & 93 & 66 & 90 & 93 & 71 & 92 & 95  \\
\mbox{Week 9} & 63 & 82 & 88 & 84 & 90 & 94 & 77 & 91 & 96  \\
\mbox{Week 10} & 61 & 87 & 92 & 75 & 93 & 96 & 74 & 94 & 98 \\
\mbox{Week 11} & 58 & 86 & 89 & 77 & 93 & 94 & 68 & 91 & 95 \\
\mbox{Week 12} & 69 & 90 & 92 & 69 & 97 & 96 & 73 & 93 & 98 \\
\mbox{Week 13} & 60 & 87 & 90 & 74 & 91 & 94 & 77 & 94 & 96 \\
\mbox{Week 14} & 67 & 87 & 89 & 81 & 92 & 95 & 69 & 89 & 94 \\
\mbox{Week 15} & 66 & 91 & 95 & 65 & 93 & 96 & 63 & 93 & 97 \\
\mbox{Week 16} & 65 & 91 & 93 & 79 & 94 & 97 & 62 & 91 & 96 \\
\mbox{Week 17} & 68 & 90 & 95 & 81 & 93 & 98 & 71 & 95 & 98 \\ \hline
\end{tabular}}
\end{table}

Comparing the results of these studies, we find that sometimes,
paradoxically, the model gives better results using only one week's
data rather than all. However, Corollary \ref{pred:var:paradox} in
Section \ref{pred:var:simplest:dlm} predicts this finding. Because
the prior for $\sigma_{1}^{2}$ is $IG(2, 0.01)$ the expectation of
$\sigma_{1}^{2}$ is $0.01,$ implying that $\sigma_{\beta}^{2} \simeq
0.01$ and $\sigma_{\delta}^{2} \simeq 0.01\times0.02.$ Hence,
$\sigma_{\beta}^{2} \left( 1 +
\frac{\sigma_{\beta}^{2}}{\sigma_{\delta}^{2}} \right) \simeq 0.51,$
which is less than $\sigma_{\varepsilon}^{2}$ (for example, the
median of $\sigma^{2}$ is around 1.21 in Study $\tilde{B}$ and even
larger in Study $\tilde{A}$). By the sufficient and necessary
condition in Corollary \ref{pred:var:paradox}, the predictive
variance of Study $\tilde{A}$ is less than that of Study
$\tilde{B}.$ However, notice that $\sigma^{2}$ and $\lambda$ vary
from week to week in $\tilde{A},$ which may also lead to the paradox
observed in the empirical findings of this section. For example, in
(b) of Figure \ref{cov:prob:paper}, the coverage probability of
$\tilde{B}$ at the $4^{\mbox{\tiny{th}}}$ week is larger than that
of $\tilde{A}.$ From the above discussion, we know that the
predictive variance of $\tilde{A}$ should be less than that of
$\tilde{B}.$ However, $\sigma^{2}$ of $\tilde{A}$ is larger than
that of $\tilde{B},$ leading an inflated predictive variance of
$\tilde{A}.$ This feature makes it difficult to compare these two
predictive variances, but explains the paradox we see in those
figures.

\section{Summary and Conclusions}\label{chapter2:summary}

To assess the dynamic linear modelling approach to modelling
space--time fields, we have applied it to an hourly ozone
concentration field over a geographical spatial domain covering most
of the eastern United States. To focus that assessment we consider
just one cluster of spatial sites we call Cluster 2 during a single
ozone season. Moreover, we have used a variant of the dynamic linear
modelling approach of Huerta et al. (2004) implemented through MCMC
sampling.

Our assessment reveals some difficulties with that very flexible
approach and practical challenges that it presents. We also have
made some recommendations on improvement.

A curious finding is the posterior dependence of $\lambda$ and
$\sigma^{2}$, in contradiction to our prior assumption. Although the
very efficient method Huerta et al. (2004) propose to sampling these
parameters is biased, that bias does not appear large enough to
account for that phenomenon. We also discovered that the assumption
of their constancy over time is untenable.

The coverage probabilities of the model's posterior predictive
credibility intervals over successive weeks, conditional on all $17$
weeks of data, increase monotonically. Counter to intuition, that
would imply more and more uncertainty as time evolves, an artifact
of the modelling that seems hard to explain. A pragmatic way around
this undesirable property involves incorporating the length of the
time span of the temporal domain $T$ into the selection of the
values of the model parameters, such as $\tau_{y}^{2},$
$\tau_{1}^{2}$ and $\tau_{2}^{2}.$ Section
\ref{pred:var:simplest:dlm} studies the correlation structure of the
simplest first--order polynomial DLM and finds reasonable conditions
to impose on those parameters.

One further Study $\tilde{D}$ tests the proposed constraints on the
data. The settings are identical with those in Study $\tilde{C}$
except that $\tau_{y}^{2},$ $\tau_{1}^{2}$ and $\tau_{2}^{2}$ are
replaced by $\tau_{y}^{2}/17,$ $\tau_{1}^{2}/17$ and
$\tau_{2}^{2}/17,$ respectively, to take account of the longer $17$
week time span of our study compared to the one week time span of
the application in Huerta et al. (2004). Figure \ref{cov:prob:paper}
compares Study $\tilde{D}$ with the others. Observe that its
coverage probabilities behave like those of Study $\tilde{A}.$ This
adjustment does seem to eliminate the undesirable property of
increasing credibility bands of Studies $\tilde{B}$ and $\tilde{C}.$

Another possible approach to dealing with the unsuitability of fixed
model parameters uses the composition of Metropolis--Hasting
kernels. In other words, we could include these parameters in the
Metropolis--Hasting algorithm as in Section \ref{ffbs:state:sample}.
We can use six Metropolis--Hasting kernels to sample from the target
distribution $\pi({\mathbf{\gamma}}|y_{1:T})$, updating each
component of $\gamma$ iteratively, where $\gamma$ has defined in
Section \ref{chapter2:dlmbackgrand}. But, not surprisingly that
approach fails because of the extreme computational burden it
entails. However, that alternative is the subject of current work
along with an approach that admits time varying $\lambda$s and
$\sigma^{2}$s.

The greatest difficulty involved in the use of the DLM in modelling
air pollution space--time fields lies in the computational burden it
entails. For that reason, we have not been able to address the
geographical domain of real interest, one that embraces $274$ sites
in the eastern United States, with 120 days of hourly ozone
concentrations. In a manuscript under preparation, an alternative
hierarchical Bayesian method that can cope with that larger domain
will be compared with the DLM where the latter can practically be
applied.
\section*{Acknowledgements}

We thank Prasad Kasibhatla of Nicholas School of the Environment of
Duke University for providing the dataset used in this paper and
helping with its installation. The funding for the work was
provided by the Natural Sciences and Engineering Research
Council of Canada.

\appendix
\section{Supplementary results}\label{appendix:chapter2:inference}

\subsection{Results for Section \ref{pred:var:simplest:dlm}}\label{pred:var:Result3}

Only the results about the predictive variances of $y_{01}|y_{11}$
and $y_{01}|y_{11}, y_{12}$ are shown in this appendix. The other
two cases can be obtained by the same method. Refer to Theorem
\ref{pred:var:simple:dlm:result}, the predictive variance of
$y_{01}|y_{11}$ can also be written as follows:
\begin{eqnarray*}
Var(y_{01}|y_{11}) & = & (1-\exp(- \frac{d_{01}}{\lambda}))
\sigma_{\varepsilon}^{2} \left\{ 2 -
\frac{1-\exp(-\frac{d_{01}}{\lambda})}{1+\frac{\sigma_{\beta}^{2}+
\sigma_{\delta}^{2}}{\sigma_{\varepsilon}^{2}}}
\right\}.
\end{eqnarray*}

The first partial derivatives of this predictive variances regarding
to $d_{01},$ $\lambda$ and $\sigma_{\varepsilon}^{2}$ are given by:
 \begin{eqnarray*}
\frac{\partial}{\partial d_{01}} Var(y_{01}|y_{11}) =
\frac{2d_{01}}{\lambda}\exp(-\frac{d_{01}}{\lambda})\sigma_{\varepsilon}^{2}\frac{
\sigma_{\beta}^{2} + \sigma_{\delta}^{2} +
\sigma_{\varepsilon}^{2}\exp(-\frac{d_{01}}{\lambda})
}{\sigma_{\beta}^{2} + \sigma_{\delta}^{2} +
\sigma_{\varepsilon}^{2} },
\end{eqnarray*}

\begin{eqnarray*}
\frac{\partial}{\partial \lambda} Var(y_{01}|y_{11}) =
-\frac{2d_{01}}{\lambda^{2}}\exp(-\frac{d_{01}}{\lambda})
\sigma_{\varepsilon}^{2}\frac{\sigma_{\beta}^{2}
+ \sigma_{\delta}^{2} +
\sigma_{\varepsilon}^{2}\exp(-\frac{d_{01}}{\lambda})
}{\sigma_{\beta}^{2} + \sigma_{\delta}^{2} +
\sigma_{\varepsilon}^{2} },
\end{eqnarray*}
and
\begin{eqnarray*}
\addtolength{\linewidth}{1cm} \frac{\partial}{\partial
\sigma_{\varepsilon}^{2}} Var(y_{01}|y_{11}) & = &
(1-\exp(-\frac{d_{01}}{\lambda})) \left\{ 2 - (1-
\exp(-\frac{d_{01}}{\lambda})) \sigma_{\varepsilon}^{2}\frac{
\sigma_{\varepsilon}^{2} + 2\sigma_{\beta}^{2} +
2\sigma_{\delta}^{2} } { ( \sigma_{\varepsilon}^{2} +
\sigma_{\beta}^{2}
+ \sigma_{\delta}^{2} )^{2} }  \right\} \\
& > & (1-\exp(-\frac{d_{01}}{\lambda}))\left\{ 2 -
\frac{ \sigma_{\varepsilon}^{2}( 2\sigma_{\beta}^{2} +
2\sigma_{\delta}^{2} + \sigma_{\varepsilon}^{2} ) }
{ ( \sigma_{\beta}^{2} + \sigma_{\delta}^{2}
+ \sigma_{\varepsilon}^{2} )^{2} }  \right\} \\
& = & \frac{ 1-\exp(-\frac{d_{01}}{\lambda}) } {(\sigma_{\beta}^{2}
+ \sigma_{\delta}^{2} + \sigma_{\varepsilon}^{2})^{2} } \{
2(\sigma_{\beta}^{2} + \sigma_{\delta}^{2} )^{2} +
\sigma_{\varepsilon}^{4} +
2\sigma_{\varepsilon}^{2}(\sigma_{\beta}^{2} + \sigma_{\delta}^{2})
\},
\end{eqnarray*}
respectively. It is straightforward to obtain that
$Var(y_{01}|y_{11})$ is increasing when $d_{01}$ increases, or
$\lambda$ decreases, or $\sigma_{\varepsilon}^{2}$ increases. We
next show these properties also hold for $Var(y_{01}|y_{11},
y_{12}).$ By Theorem \ref{pred:var:simple:dlm:result},
$Var(y_{01}|y_{11}, y_{12})$ can also be written as:
$$
Var(y_{01}|y_{11}, y_{12}) =
(1-\exp(-\frac{d_{01}}{\lambda}))\sigma_{\varepsilon}^{2}\left\{  2
- \frac{ 1- \exp(-\frac{d_{01}}{\lambda}) }{ 1 + \frac{
(\sigma_{\beta}^{2} + \sigma_{\delta}^{2} )( \sigma_{\delta}^{2} +
\sigma_{\varepsilon}^{2} ) }{
\sigma_{\varepsilon}^{2}(\sigma_{\beta}^{2} + 2\sigma_{\delta}^{2} +
\sigma_{\varepsilon}^{2}) } } \right\}.
$$
The corresponding first partial derivatives are given as follows:
\begin{eqnarray*}
\frac{\partial}{\partial d_{01}} Var(y_{01}|y_{11}, y_{12}) & = &
\frac{2}{\lambda}\exp(-\frac{d_{01}}{\lambda})\sigma_{\varepsilon}^{2}\frac{A
+ \exp(-\frac{d_{01}}{\lambda})}{1 + A},
\end{eqnarray*}

\begin{eqnarray*}
\frac{\partial}{\partial \lambda} Var(y_{01}|y_{11}, y_{12}) & = &
-\frac{2d_{01}}{\lambda^{2}}\exp(-\frac{d_{01}}{\lambda})
\sigma_{\varepsilon}^{2}\frac{A + \exp(-\frac{d_{01}}{\lambda})}{1 + A},
\end{eqnarray*}
and
\begin{eqnarray*}
\frac{\partial}{\partial \sigma_{\varepsilon}^{2}}
Var(y_{01}|y_{11}, y_{12}) & = & (1-\exp(-\frac{d_{01}}{\lambda}))
\left\{ 2 - (1-\exp(-\frac{d_{01}}{\lambda}))
\frac{\sigma_{\varepsilon}^{2}}{A^{2}} ( c_{1} A - c_{2}c_{3}) \right\} \\
& > & \frac{1-\exp(-\frac{d_{01}}{\lambda})}{A^{2}}c_{4},
\end{eqnarray*}
respectively, where $A = \frac{ (\sigma_{\beta}^{2} +
\sigma_{\delta}^{2} )( \sigma_{\delta}^{2} +
\sigma_{\varepsilon}^{2} ) }{ \sigma_{\varepsilon}^{2}(
\sigma_{\beta}^{2} + 2\sigma_{\delta}^{2} +
\sigma_{\varepsilon}^{2}) },$ $c_{1} = \sigma_{\beta}^{2} +
2\sigma_{\delta}^{2} + \sigma_{\varepsilon}^{2},$ $c_{2} =
\sigma_{\beta}^{2} + \sigma_{\delta}^{2},$
$c_{3}=\sigma_{\delta}^{2}c_{1} +
\sigma_{\varepsilon}^{2}(\sigma_{\delta}^{2} +
\sigma_{\varepsilon}^{2}),$ and $c_{4} =
\sigma_{\varepsilon}^{2}c_{1}(2\sigma_{\beta}^{2} +
3\sigma_{\delta}^{2} + \sigma_{\varepsilon}^{2}) +
\sigma_{\varepsilon}^{2}c_{2}(\sigma_{\delta}^{2} +
\sigma_{\varepsilon}^{2})(3\sigma_{\beta}^{2} + 6\sigma_{\delta}^{2}
+ 4\sigma_{\varepsilon}^{2}) + c_{2}^{2}(\sigma_{\delta}^{2} +
\sigma_{\varepsilon}^{2})^{2}.$

\subsection{Results for Section \ref{chapter2:algo:est:mcmc}}
\label{appendix:section:2:3:1}

The joint posterior distribution for $x_{1:T}, \lambda$ and $\sigma^{2}$ is given by
\begin{eqnarray*}
p(x_{1:T}, \lambda, \sigma^{2} | y_{1:T}) & = & p(\lambda, \sigma^{2})p( {\bf x_{T}}|\lambda, \sigma^{2}, y_{1:T}) \prod_{t=1}^{T}p( {\bf x_{T-t}}|{\bf x_{T-t+1} }, \lambda,\sigma^{2}, y_{1:T})\\
&   & \mbox{}\prod_{t=1}^{T}p({\bf y_{t}}|\lambda, \sigma^{2}, y_{1:t-1})\\
& = & p(x_{1:T}|\lambda, \sigma^{2},
y_{1:T})p(\sigma^{2}|\lambda, y_{1:T}) p(\lambda|{\bf y_{T}}).
\end{eqnarray*}
Suppose $p(\lambda, \sigma^{2}) = p(\lambda)p(\sigma^{2}),$ that is,
the priors for $\lambda$ and $\sigma^{2}$ are independent of other.

The joint posterior distribution for $\lambda$ and $\sigma^{2}$ can
be written as follows:
\begin{eqnarray*}
p(\lambda, \sigma^{2}|y_{1:T}) & \propto & p(\lambda)p(\sigma^{2})
(\sigma^{2})^{-nT/2}\prod_{t=1}^{T}|Q_{t}|^{-1/2} \exp\left\{
-\frac{1}{2\sigma^{2}} \sum_{t=1}^{T} {\bf
e_{t}}^{\prime}Q_{t}^{-1}{\bf e_{t}}\right\}.
\end{eqnarray*}

If the prior for $\sigma^{2}$ is an inverse gamma distribution with
shape parameter $\alpha$ and scale parameter $\beta,$ then the
posterior distribution for $\sigma^{2}$ is also an inverse gamma
distribution with shape parameter $\alpha + \frac{nT}{2}$ and scale
parameter $\beta + \frac{1}{2}\sum_{t=1}^{T}{\bf
e_{t}}^{\prime}Q_{t}^{-1}{\bf e_{t}}.$

Hence, the posterior density for $\lambda$ can be written as
follows:
\begin{eqnarray*}
p(\lambda | y_{1:T}) & =       & \frac{ p(\lambda, \sigma^{2} |
y_{1:T}) } { p(\sigma^{2} | \lambda, y_{1:T})  } \\
& \propto & p(\lambda) \prod_{t=1}^{T}|Q_{t}|^{-1/2} \left[ \beta +
\frac{1}{2}\sum_{t=1}^{T}{\bf e_{t}}^{\prime}Q_{t}^{-1}{\bf e_{t}}
\right]^{-(\alpha+nT/2)}.
\end{eqnarray*}

Therefore, the posterior density for $\mathbf{x_{1:T}}$ is given by
\begin{eqnarray*}
p( x_{1:T}| \lambda, \sigma^{2}, y_{1:T}) & = & p({\bf x_{T}}
|\lambda, \sigma^{2}, y_{1:T}) \prod_{t=1}^{T} p({\bf x_{T-t}} |
{\bf x_{T-t+1}}, \lambda, \sigma^{2}, y_{1:T}).
\end{eqnarray*}

\subsection{Results for Section \ref{chapter2:algo:pred}}\label{appendix:spatial}

Given the values of the phase parameters, range and variance
parameters and the observations until time $t$, the joint
distribution of $\alpha_{1t}^{s}, {\mathbf{\alpha_{1t}}}$ is $$
\left( \begin{array}{c}
       \alpha_{t}^{s}\\
       {\bf \alpha_{1t}}
       \end{array}
\right)
\sim {\mbox{\large{N}}}\left[ \left( \begin{array}{c}
                                    \alpha_{1,t-1}^{s}\\
                                    {\bf \alpha_{1, t-1}}
                                    \end{array}
                              \right),
                              \sigma^{2}\tau_{1}^{2}\Sigma^{*}(\lambda_{1}),
                       \right]
$$
where $$
\Sigma^{*}(\theta) = \exp\{ - V^{*}/\theta\}
                   = \left[\begin{array}{ccc}
                           \Sigma_{11}^{*}(\theta) & \Sigma_{12}^{*}(\theta)\\
                           \Sigma_{21}^{*}(\theta) & \Sigma_{22}^{*}(\theta)
                           \end{array}
                     \right],
$$
with $\Sigma_{11}^{*}(\theta)$ a scalar, $\Sigma_{12}^{*}(\theta)$ a
$1$ by $n$ vector, and $\Sigma_{22}^{*}(\theta)$ a $n$ by $n$
matrix. We use $V^{*}$ to denote the new distance matrix for the
unknown site $s$ and the monitoring stations $s_{1},\ldots, s_{n}.$

We then have the conditional posterior distribution of
$\alpha_{1t}^{s}$ as follows:
\begin{equation}
\begin{array}{lcl}
(\alpha_{1t}^{s} | \alpha_{1,t-1}^{s}, {\bf \alpha_{1t}}, {\bf
\alpha_{1,t-1}}, {\bf y_{t}}, \lambda, \sigma^{2}) & \sim & N[
\alpha_{1,t-1}^{s} + \Sigma^{*}_{12}(\lambda_{1})\Sigma^{*}_{22}
(\lambda_{1})^{-1}({\bf \alpha_{1t}} - {\bf \alpha_{1,t-1}}), \\
& &  \sigma^{2}\tau_{1}^{2}( \Sigma^{*}_{11} (\lambda_{1})  -
\Sigma^{*}_{12}(\lambda_{1})
\Sigma^{*}_{22}(\lambda_{1})^{-1}\Sigma^{*}_{21}(\lambda_{1})) ].
\end{array}
\label{alpha1t:pred:post}
\end{equation}

Similarly, the conditional posterior distribution for
$\alpha_{2t}^{s}$ is
\begin{equation}
\begin{array}{lcl}
(\alpha_{2t}^{s} | \alpha_{2,t-1}^{s}, {\bf \alpha_{2t}}, {\bf
\alpha_{2,t-1}}, {\bf y_{t}}, \lambda, \sigma^{2}) & \sim & N[
\alpha_{2,t-1}^{s} + \Sigma^{*}_{12}(\lambda_{2})
\Sigma^{*}_{22}(\lambda_{2})^{-1}({\bf \alpha_{2t}} - {\bf
\alpha_{2,t-1}}), \\
& & \sigma^{2}\tau_{2}^{2}( \Sigma^{*}_{11} (\lambda_{2})  -
\Sigma^{*}_{12}(\lambda_{2})\Sigma^{*}_{22}(\lambda_{2})^{-1}\Sigma^{*}_{21}
(\lambda_{2})) ].
\end{array}
\label{alpha2t:pred:post}
\end{equation}

Using the observation equation as in Model (\ref{dlm:y}), we have
the conditional predictive distribution for $y_{t}^{s}$ as follows:
\begin{equation}
\begin{array}{lcl}
(y_{t}^{s} | {\bf y_{t}}, \alpha_{1t}^{s}, \alpha_{2t}^{s}, {\bf
\alpha_{1t}}, {\bf \alpha_{2t}}, \beta_{t}, \lambda, \sigma^{2})
& \sim & N[ \beta_{t} + S_{1t}(a_{1})\alpha_{1t}^{s} + S_{2t}(a_{2})\alpha_{2t}^{s} \\
&      & + \Sigma^{*}_{12}(\lambda)\Sigma^{*}_{22}(\lambda)^{-1}({\bf y_{t}} - {\mathbf{1}_{n}}\beta_{t} \\
&      & - S_{1t}(a_{1}){\bf \alpha_{1t}} - S_{2t}(a_{2}){\bf \alpha_{2t}}),\\
&      & \sigma^{2}(\Sigma^{*}_{11}(\lambda) -
\Sigma^{*}_{12}(\lambda)\Sigma^{*}_{22}(\lambda)^{-1}\Sigma^{*}_{21}(\lambda))].
\end{array}
\label{yt:pred:post}
\end{equation}

The software is in http://enviro.stat.ubc.ca.


\end{document}